\newcommand{\bea}{\begin{eqnarray}}
\newcommand{\eea}{\end{eqnarray}}
\newcommand{\be}{\begin{equation}}
\newcommand{\ee}{\end{equation}}
\newcommand{\ben}{\begin{enumerate}}
\newcommand{\een}{\end{enumerate}}
\newcommand{\bi}{\begin{itemize}}
\newcommand{\ei}{\end{itemize}}
\newcommand{\bmi}[1]{\begin{minipage}{#1 cm}}
\newcommand{\emi}{\end{minipage}}

\newcommand{\rund}[1]{\left(#1\right)}
\newcommand{\vc}[1]{\mbox{\boldmath $#1$}}
\renewcommand{\d}{{\rm d}}

\newcommand{\eck}[1]{\left[ #1 \right]}
\newcommand{\ave}[1]{\left\langle #1 \right\rangle}
\newcommand{\vt}{\vartheta}
\newcommand{\vp}{\varphi}
\newcommand{\eps}{\epsilon}

\def\llabel#1{\label{sc:#1}}
\def\elabel#1{\label{eq:#1}}
\def\flabel#1{\label{fig:#1}}

{\catcode`\@=11
\gdef\SchlangeUnter#1#2{\lower2pt\vbox{\baselineskip 0pt \lineskip0pt
  \ialign{$\m@th#1\hfil##\hfil$\crcr#2\crcr\sim\crcr}}}
}

\def\lesssim{\mathrel{\mathpalette\SchlangeUnter<}}

\newcommand{\om}{\Omega_\mr m}
\newcommand{\sig}{\sigma_8}

\newcommand{\vpi}{\vec \pi}

\newcommand{\mr}{\mathrm}
\newcommand{\tbf}{\textbf}

\newcommand{\tn}{\textnormal}

\newcommand{\beq}{\begin{eqnarray}}
\newcommand{\eeq}{\end{eqnarray}}

\documentclass[]{aa}
\usepackage{graphicx}
\usepackage{epsfig}
\usepackage{txfonts}
\usepackage{color}
\usepackage{natbib}
\usepackage{subfigure} 
\bibpunct{(}{)}{;}{a}{}{,}
\begin{document}
\title{COSEBIs: Extracting the full E-/B-mode information from cosmic shear correlation functions}

   \author{Peter Schneider\inst{1} \and Tim Eifler\inst{2,1} \and Elisabeth Krause\inst{3}}

\offprints{Peter Schneider}
\institute{Argelander-Institut f\"ur Astronomie, 
  Universit\"at Bonn, Auf dem H\"ugel 71,
  D-53121 Bonn, Germany, \email{peter@astro.uni-bonn.de}\and 
Center for Cosmology and Astro-Particle Physics, The Ohio State University,
191 W. Woodruff Ave., Columbus, OH 43210, USA,
\email{teifler@mps.ohio-state.edu} 
\and 
California Institute of Technology, Dept.\ of Astronomy,
MC 105-24, Pasadena CA 91125, USA, \email{ekrause@astro.caltech.edu}}

\titlerunning{COSEBIs: Extracting the full E-/B-mode information from
  cosmic shear correlation functions} 

   \date{Received ; accepted }

   \abstract{Cosmic shear is considered one of the most powerful
     methods for studying the properties of Dark Energy in the
     Universe. As a standard method, the two-point correlation
     functions $\xi_\pm(\vt)$ of the cosmic shear field are used as
     statistical measures for the shear field.}{In order to separate
     the observed shear into E- and B-modes, the latter being most
     likely produced by remaining systematics in the data set and/or
     intrinsic alignment effects, several statistics have been defined
     before. Here we aim at a complete E-/B-mode decomposition of the
     cosmic shear information contained in the $\xi_\pm$ on a finite
     angular interval.}{We construct two sets of such E-/B-mode
     measures, namely Complete Orthogonal Sets of E-/B-mode Integrals
     (COSEBIs), characterized by weight functions between the
     $\xi_\pm$ and the COSEBIs which are polynomials in $\vt$ or
     polynomials in $\ln\vt$, respectively. Considering the likelihood
     in cosmological parameter space, constructed from the COSEBIs, we
     study their information content.}{We show that the information
     grows with the number of COSEBI modes taken into account, and
     that an asymptotic limit is reached which defines the maximum
     available information in the E-mode component of the
     $\xi_\pm$. We show that this limit is reached the earlier (i.e.,
     for a smaller number of modes considered) the narrower the
     angular range is over which $\xi_\pm$ are measured, and it is
     reached much earlier for logarithmic weight functions. For
     example, for $\xi_\pm$ on the interval $1'\le \vt\le 400'$, the
     asymptotic limit for the parameter pair $(\Omega_{\rm
       m},\sigma_8)$ is reached for $\sim 25$ modes in the linear
     case, but already for 5 modes in the logarithmic case. The
     COSEBIs form a natural discrete set of quantities, which we
     suggest as method of choice in future cosmic shear likelihood
     analyses.}{}
   
   \maketitle
%

\section{\llabel{1}Introduction}
\def\tmin{{\vt_{\rm min}}}
\def\tmax{{\vt_{\rm max}}}

The shear field in weak lensing is caused by the tidal component of
the gravitational field of the mass distribution between us and a
distant population of sources \citep[see][for recent
reviews]{mel99,bas01,ref03,skw06,mvv08}. If the shear, estimated from
the image shapes of distant galaxies, is solely due to gravitational
lensing, then it should consist only of a `gradient component', the
so-called E-mode shear \citep[see][]{cnp02,svm02}. B-modes (or curl
components) cannot be generated by gravitational light deflection in
leading order, and higher-order effects from lensing are expected to
be small, as seen in ray-tracing simulations through the cosmological
density field \citep[e.g.,][]{jsw00,hhw09}.

Therefore, the splitting of the observered shear field into its E- and
B-modes is of great importance to isolate the gravitational shear from
the shear components most likely not due to lensing, in order to (i)
have a measure for the impact of other effects besides lensing (such
as insufficient PSF correction for the shape measurements, or
intrinsic alignment effects) on the observed shear field, and to (ii)
isolate the lensing shear and to compare it with the expectation from
cosmological models. Indeed, almost all more recent cosmic shear
surveys perform such an E-/B-mode decomposition of second-order shear
measures \citep[e.g.,][]{hyg02,jbf03,hssh07,fsh08}.

The standard technique for this separation is the aperture dispersion
$\ave{M_{\rm ap}^2(\theta)}$ and $\ave{M_\times^2(\theta)}$
\citep{swj98}, which can be calculated in terms of the shear two-point
correlation functions (2PCFs) $\xi_\pm(\vt)$ on a finite interval
$0\le \vt\le 2\theta$. Alternatively, one can construct E- and B-mode
shear correlation functions \citep{cnp02}, which, however, can be
calculated only if the shear correlation function $\xi_+$ is known for
arbitrarily large separations. As was pointed out by \cite{kse06}, the
fact that the calculation of the aperture dispersion requires the
knowledge of the shear correlation functions down to zero separation,
together with the inability to measure the shape of image pairs with
very small angular separation, leads to biases in the estimated values
for the aperture dispersions, in particular to an effective E-/B-mode
mixing.

For that reason, \cite{sck07} -- hereafter SK07 -- developed a new
second-order shear statistics, that can be calculated from the shear
correlation functions $\xi_\pm$ on a finite interval $\tmin\le
\vt\le\tmax$ and which provides a clean separation of E- and
B-modes. In particular, SK07 derived general expressions for the
relation between E-/B-mode second-order shear quantities and the shear
2PCFs. They considered one particular example of such a relation,
leading to the so-called the ring statistics, based solely on
geometric considerations. \cite{esk09} and \cite{fuk10} -- hereafter
FK10 -- have shown that, although the signal-to-noise at fixed angular
scale is smaller for the ring statistics than for the aperture
dispersion, the correlation matrix between measurements at different
angular scales is considerably narrower in the case of the ring
statistics, yielding that the information contents of the two measures
are quite comparable. Applying the ring statistics to the same cosmic
shear correlation functions as used by \cite{fsh08} in their
measurement from the Canada-France-Hawaii Telescope Legacy Survey,
\cite{esk09} obtained a
clear signal, as well as a better localization of the
remaining B-modes.

In FK10, more general E-/B-mode measures have been considered, based
on the general transformation derived in SK07. Specifically, FK10 have
constructed E-mode quantities which maximize the signal-to-noise for a
given interval $\tmin\le \vt\le\tmax$, or which maximize the figure of
merit in parameter space, as obtained from considering the Fisher
matrix. Both of the resulting E-mode statistics are by construction
superior to the ring statistics, and also yield higher
signal-to-noise, or a higher figure-of-merit, than the aperture
dispersion. 

In this paper, we construct sets of E-/B-mode measures, $E_n$ and
$B_n$, based on shear correlation functions on a finite interval. In a
well-defined sense, for a given angular interval $\tmin\le \vt\le
\tmax$, these second-order E-/B-mode measures form a complete set
each, so that all E-B-separable information contained in the
$\xi_\pm(\vt)$ is also contained in this complete set. With these
complete sets of second-order shear measures, we propose a new
approach to compare observed shear correlations with model
predictions. Whereas all such comparisons done hitherto define a
second-order shear measure as a function of angular scale [such as
$\xi_\pm(\vt)$ or $\ave{M_{\rm ap}^2(\theta)}$], the choice of the
grid points in the angular scale being arbitrary, the complete set of
the $E_n$ are a `natural' discrete set of quantities that can be used
in a likelihood analysis. One can hope that a finite and possibly
rather small number of the $E_n$ contains most of the cosmological
information, depending on the choice of the set.

In Sect.\ts\ref{sc:EB-decomp} we summarize the general equations for
E-/B-mode measures obtained from the two-point correlation functions
of the shear field over a finite interval, and derive the covariance
matrix for a set of such E-B-mode measures. We then construct in
Sect.\ts\ref{sc:COSEB} two examples of Complete Orthogonal Sets of
E-/B-mode Integrals (COSEBIs), one of them using weight functions
which are polynomials in $\vt$, the others being polynomials in
$\ln\vt$. In the former case, explicit relations for the corresponding
weight functions are obtained for any polynomial order, whereas in the
logarithmic case the coefficients have to be obtained through a matrix
inversion. In Sect.\ts\ref{sc:likelihoodanalysis}, we then
investigate the information content of these COSEBIs, by calculating
the likelihood of cosmological parameter combinations and the
corresponding Fisher matrix for a fiducial cosmic shear survey, using
the two COSEBIs constructed, as well as the original shear correlation
functions. We conclude by discussing the advantages of the COSEBIs
over the other second-order shear measures that have been suggested in
the literature.  In Appendix\ts\ref{sc:SN-maxi}, we show how COSEBIs
can be used to maximize the signal-to-noise of a cosmic shear E-mode
measure. In addition we show how to construct pure E/B-mode
correlation functions from the COSEBIs and relate them to the 2PCF.

\section{\llabel{EB-decomp}E-/B-mode decomposition}
In SK07 we have shown than an E-/B-mode separation of second-order
shear statistics is  obtained from the 2PCFs
$\xi_\pm$ by
\bea
\label{eq:EBmodes}
E&=&{1\over 2}\int_0^\infty \d\vt\;\vt\,\eck{T_+(\vt)\xi_+(\vt)
+T_-(\vt)\xi_-(\vt)} \;,\nonumber \\
B&=&{1\over 2}\int_0^\infty \d\vt\;\vt\,\eck{T_+(\vt)\xi_+(\vt)
-T_-(\vt)\xi_-(\vt)} \;,
\eea
provided the two weight functions $T_\pm$ are related through
\be
\int_0^\infty\d\vt\,\vt\,{\rm J}_0(\ell\vt) T_+(\vt) 
=\int_0^\infty\d\vt\,\vt\,{\rm
  J}_4(\ell\vt) T_-(\vt)
\elabel{TpmJ04}
\ee
or, equivalently,
\bea
T_+(\vt)&=&T_-(\vt)+\int_\vt^\infty\d\theta\;\theta\,T_-(\theta)
\rund{{4\over\theta^2}-{12\vt^2\over \theta^4}}\;, \nonumber \\
T_-(\vt)&=&T_+(\vt)+\int_0^\vt\d\theta\;\theta\,T_+(\theta)
\rund{ {4\over \vt^2}-{12\theta^2\over \vt^4}}\;.
\elabel{Tplusminus}
\eea
In this case, $E$ contains only E-modes, whereas $B$ depends only on
the B-mode shear. Furthermore, it was shown in SK07 that an E-mode
second-order statistics is obtained from the shear correlation
functions on a finite interval $\tmin\le \vt\le \tmax$ if the function
$T_+$ vanishes outside the same interval, and in addition, the two
conditions 
\be
\int_{\tmin}^{\tmax}\d\vt\;\vt\,T_+(\vt)=0=
\int_{\tmin}^{\tmax}\d\vt\;\vt^3\,T_+(\vt)
\elabel{conditions}
\ee
are satisfied; in this case, the function $T_-(\vt)$ as calculated
from Eq.\ts(\ref{eq:Tplusminus}) also has finite support on the interval
$\tmin\le \vt\le \tmax$. In SK07, a particular set of functions
$T_\pm$ was introduced, originating from the geometrical construction
of cross-correlating the shear in two non-overlapping annuli, and the
corresponding estimators were termed `ring statistics'. 

The origin of the conditions expressed in Eq.\ts(\ref{eq:conditions})
can be understood as follows: A uniform shear field cannot be assigned
an E- or B-mode origin. Such a shear field gives rise to shear
correlation functions of the form $\xi_+(\vt)={\rm const.}$ and
$\xi_-(\vt)=0$. According to the first of Eq.\ts(\ref{eq:conditions}),
this component is filtered out in Eq.\ts(\ref{eq:EBmodes}). Furthermore,
one possibility to distinguish between E- and B-modes is the
consideration of the vector field $\vc
u=(\gamma_{1,1}+\gamma_{2,2},\gamma_{2,1}-\gamma_{1,2})$ constructed
from partial derivatives of the shear field $\gamma(\vc\vt)$
\citep{kais95}. A pure E-mode shear yields a vanishing curl of $\vc
u$, whereas a pure B-mode shear leads to $\nabla \cdot \vc u=0$; a
shear field which yields $\nabla \cdot \vc u=0={\rm curl}(\vc u)$
cannot be uniquely classified as E- or B-mode.  

If we now consider a shear field which depends linearly on $\vc\vt$,
then the vector field $\vc u$ is constant, and thus it cannot be
uniquely split into E- and B-modes. On the other hand, such a shear
field gives rise to correlation functions of the form
$\xi_+(\vt)=A+B\vt^2$, $\xi_-(\vt)=0$, where $A$ and $B$ are
constants. Again, the correlation function of such a shear field is
filtered out due to the conditions in Eq.\ts(\ref{eq:conditions}).

\subsection{E-/B-modes from a set of functions}
Of course, there are many functions $T_+(\vt)$ which satisfy the
constraints in Eq. (\ref{eq:conditions}). Assume we construct a set of
functions $T_{+n}(\vt)$ which all satisfy Eq. (\ref{eq:conditions})
and which are, in a way specified later, orthogonal. Then one can
construct the corresponding \tbf{$T_{-n}(\vt)$} from
Eq. (\ref{eq:Tplusminus}), and thus one obtains the set $E_n$ and
$B_n$ of second-order shear measures with a clean E-/B-mode
separation. Each of the $E_n$ and $B_n$ measures an integral over the
power spectrum of E- and B-modes, respectively,
\bea
E_n&=&\int_0^\infty {\d\ell\;\ell\over 2\pi} P_{\rm E}(\ell)\,W_n(\ell)
\;,\nonumber\\
B_n&=&\int_0^\infty {\d\ell\;\ell\over 2\pi} P_{\rm B}(\ell)\,W_n(\ell) \;,
\elabel{EBfromP}
\eea
where the filter functions are
\be
W_n(\ell)=\int_{\tmin}^{\tmax}\d\vt\;\vt\,
T_{+n}(\vt)\,{\rm J}_0(\ell\vt)\;,
\elabel{WnFilter}
\ee
and where we made use of the relation between the shear correlation
functions and the power spectra \citep[see, e.g., ][]{svm02}
\bea
\xi_+(\vt)&=&\int_0^\infty {\d\ell\;\ell\over 2\pi}\,{\rm J}_0(\ell\vt)
\eck{P_{\rm E}(\ell)+P_{\rm B}(\ell)} \;,\nonumber \\
\xi_-(\vt)&=&\int_0^\infty {\d\ell\;\ell\over 2\pi}\,{\rm J}_4(\ell\vt)
\eck{P_{\rm E}(\ell)-P_{\rm B}(\ell)} \;.
\elabel{xi+-}
\eea
We next calculate the covariance of the E- and B-mode measures making
use of Eq. (\ref{eq:EBfromP}),
\bea
C^{\rm E}_{mn}&\equiv& \ave{E_m E_n}-\ave{E_m}\ave{E_n} \nonumber \\
&=&\int_0^\infty {\d\ell\;\ell\over 2\pi} W_m(\ell)
\int_0^\infty {\d\ell'\;\ell'\over 2\pi} W_n(\ell')
\ave{\Delta P_{\rm E}(\ell)\,\Delta P_{\rm E}(\ell')} \nonumber \\
&=&{1\over \pi A} \int_0^\infty \d\ell\;\ell\,W_m(\ell) W_n(\ell)
\eck{P_{\rm E}(\ell)+N_\eps}^2\;, 
\elabel{CEmn}
\eea
where in the final step we have assumed a Gaussian shear field and
used the corresponding expression for the covariance of the power
spectrum from \cite{jse08}. Here, $A$ is the survey area,
$N_\eps=\sigma_\eps^2/(2 \bar n)$ is the amplitude of the white noise
power spectrum resulting from the intrinsic ellipticity distribution
of sources, $\sigma_\eps$ is the dispersion of the intrinsic ellipticity,
and $\bar n$ is the mean number density of sources. The covariance of
the $B_n$, $C^{\rm B}_{mn}$, has exactly the same form, with $P_{\rm
  E}$ replaced by $P_{\rm B}$, and the covariance between the $E_n$
and $B_m$ vanishes.

As a consistency check, we calculate the covariance in a different
form, starting from the relation between the $E_n$ and the shear
correlation functions. We then obtain
\bea
C^{\rm E}_{mn}&=&{1\over 4}\int_{\tmin}^{\tmax}\d\vt\;\vt
\int_{\tmin}^{\tmax}\d\vt'\;\vt' \nonumber \\
&\times&\sum_{\mu,\nu=\{+,-\} }
T_{\mu m}(\vt)\, T_{\nu n}(\vt')\, C_{\mu\nu}(\vt,\vt')\;,
\elabel{CE2PCF}
\eea
where $C_{\pm\pm}(\vt,\vt')$ is the covariance of the shear
correlation function $\xi_\pm(\vt)$. Using the relations of
\cite{jse08} for the covariance of the $\xi_\pm$, assuming a Gaussian
shear field, and making use of Eq.\ts(\ref{eq:TpmJ04}), the result
(\ref{eq:CEmn}) is re-obtained.

The comparison of the $E_n^{\rm obs}$ obtained from observations
with those of a model $E_n(\vc\pi)$, where $\vc\pi$ denotes a set of
$M$ model parameters, can then be done via
\be
\chi^2= \sum_{m,n=1}^N \eck{E_m^{\rm obs}-E_m(\vc \pi)} 
\rund{C^{\rm E}}^{-1}_{mn}\eck{E_n^{\rm obs}-E_n(\vc \pi)}\;,
\ee
where $N$ is the maximum number of E-modes considered, or with a
likelihood function
\be
{\cal L}=\eck{(2\pi)^{N/2}\,\sqrt{\det C^{\rm E}}}^{-1}\,
{\rm e}^{-\chi^2/2} \;. 
\ee

\subsection{Calculation of E-mode second-order statistics from
  ray-tracing simulations} 
Due to the limited range of validity of analytic approximations for
the calculation of cosmic shear statistics, ray tracing through N-body
simulated three-dimensional density distributions are carried out
\citep[see, e.g.,][and references therein]{jsw00,hhw09}. As shown in
these papers, the resulting B-mode shear is several orders of
magnitude smaller than the E-mode shear, so that the resulting shear
field can be described very accurately in terms of an equivalent
surface mass density $\kappa(\vc\theta)$. It is often faster to derive
statistical properties of the resulting shear field from the
corresponding properties of the $\kappa$-field. For example, the
aperture mass $M_{\rm ap}$ \citep[][]{sch96} can be obtained from the
shear field through a radial filter function $Q$, but also from
the $\kappa$-field through a related radial filter function
$U$. Hence, one can calculate the field of $M_{\rm ap}$ from the
equivalent surface mass density, convolved with the filter $U$, and
the aperture mass dispersion is then given as the dispersion of this
field. In this way, no correlation functions of the shear need to be
obtained for making predictions, saving computation time.

Here we will show that, similar to the case of the aperture mass
  dispersion, the E-mode second-order shear statistics defined in
  Eq.\ts(\ref{eq:EBmodes}) can be obtained from a simulated
  $\kappa$-field, without the need to calculate the shear correlation
  functions. For that we note that,
in the absence of B-modes, one has
\[
E=\int_0^\infty \d\vt\;\vt\,T_+(\vt)\xi_+(\vt)\;,
\]
and that the correlation functions of $\kappa$ and $\gamma$ agree,
\[
\ave{\kappa(\vc\theta)\kappa(\vc\theta')}
=\ave{\gamma(\vc\theta)\gamma^*(\vc\theta')}
=\xi_+(|\vc\theta-\vc\theta'|)\;.
\]
If we smooth the convergence field with a radial filter function $F$,
obtaining 
\be
\kappa_{\rm s}(\vc\theta)=\int \d^2\theta'\; \kappa(\vc\theta')\,
F(|\vc\theta-\vc\theta'|)\;,
\ee
the correlator of the smoothed field with the unsmoothed field 
  at zero lag becomes
\be
\ave{\kappa(\vc\theta)\,\kappa_{\rm s}(\vc\theta)}
=\int \d^2\theta'\;
F(|\vc\theta-\vc\theta'|)\,\xi_+(|\vc\theta-\vc\theta'|)
\;.
\ee
Setting $\vc\vt=\vc\theta'-\vc\theta$, we see that
\be
E=\ave{\kappa(\vc\theta)\,\kappa_{\rm s}(\vc\theta)}\;,
\ee
if we choose $F(\vt)=(2\pi)^{-1} T_+(\vt)$. Hence, the calculation of
$E$ from simulations can proceed by convolving the $\kappa$-field with
the function $T_+(\vt)/(2\pi)$, and correlating the resulting field
with the original $\kappa$-field, dropping a band of width $\tmax$
along the boundaries of the field where the convolution via FFT causes
artifacts.

\section{\llabel{COSEB}Complete sets of weight functions}
Here, we construct complete sets of functions which satisfy the
constraints (\ref{eq:conditions}) for the weight function
$T_+(\vt)$ on the interval $\tmin\le\vt\le\tmax$. It should be noted
that, once a complete set of such functions is known, the maximization
of the signal-to-noise of the second-order E-mode shear -- a problem
considered in FK10 -- reduces to a linear algebra problem, as shown in
Appendix\ts\ref{sc:SN-maxi}. 

Readers less interested in the explicit construction of these COSEBIs
can go directly to Sect.\ts\ref{sc:likelihoodanalysis}.

\subsection{Polynomial weight functions}
\llabel{polynomials}

\begin{figure}
\includegraphics[width=9cm]{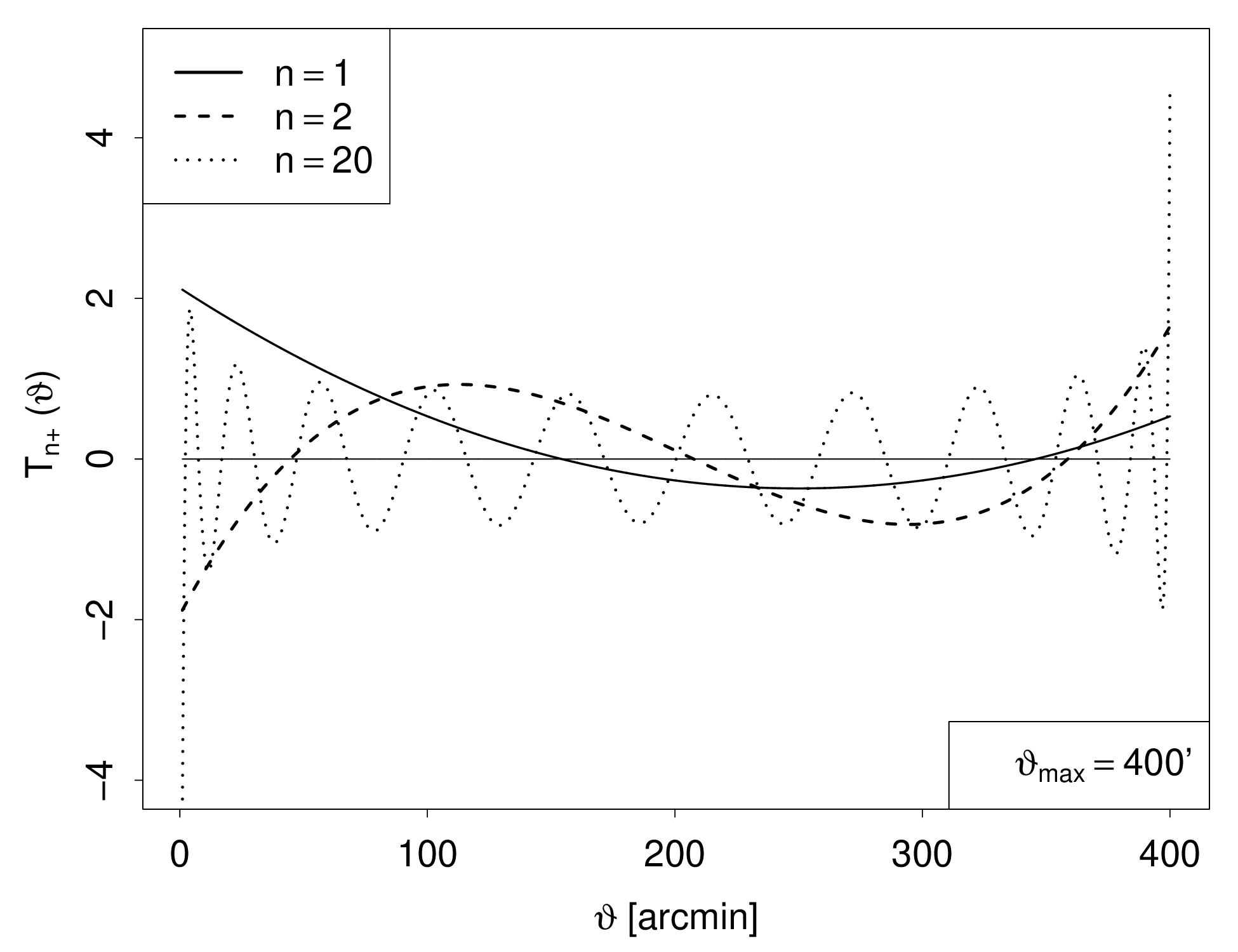}
\includegraphics[width=9cm]{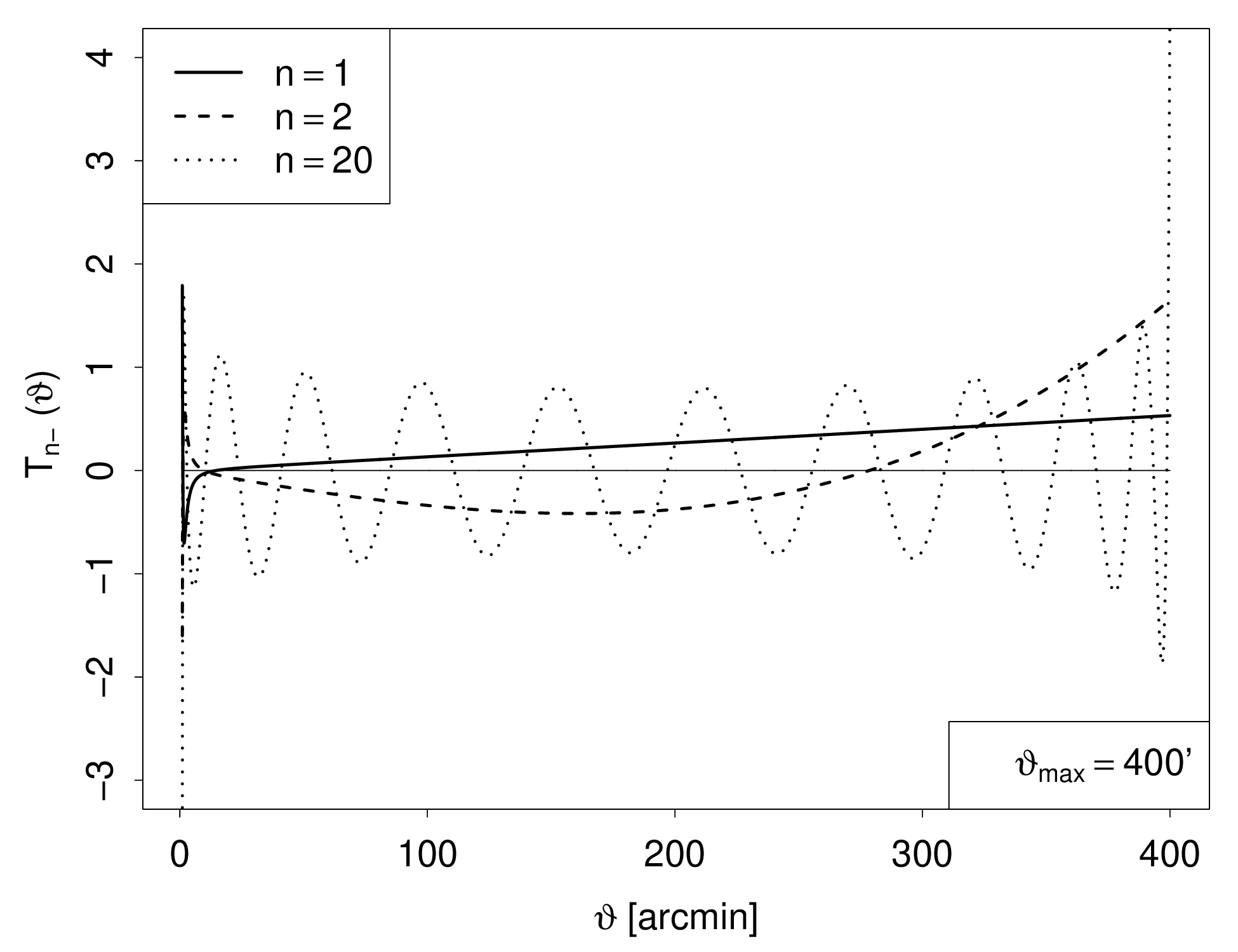}
  \caption{The linear filter functions $T_{\pm n}(\vt)$ for $\vt_{\rm
      min}=1'$, $\vartheta_\mr{max}=400'$. Note that the shape of the
    curves depends only on the ratio $\tmin/\tmax$ }
         \label{fig:T_lin}
\end{figure}
\begin{figure}
\includegraphics[width=9cm]{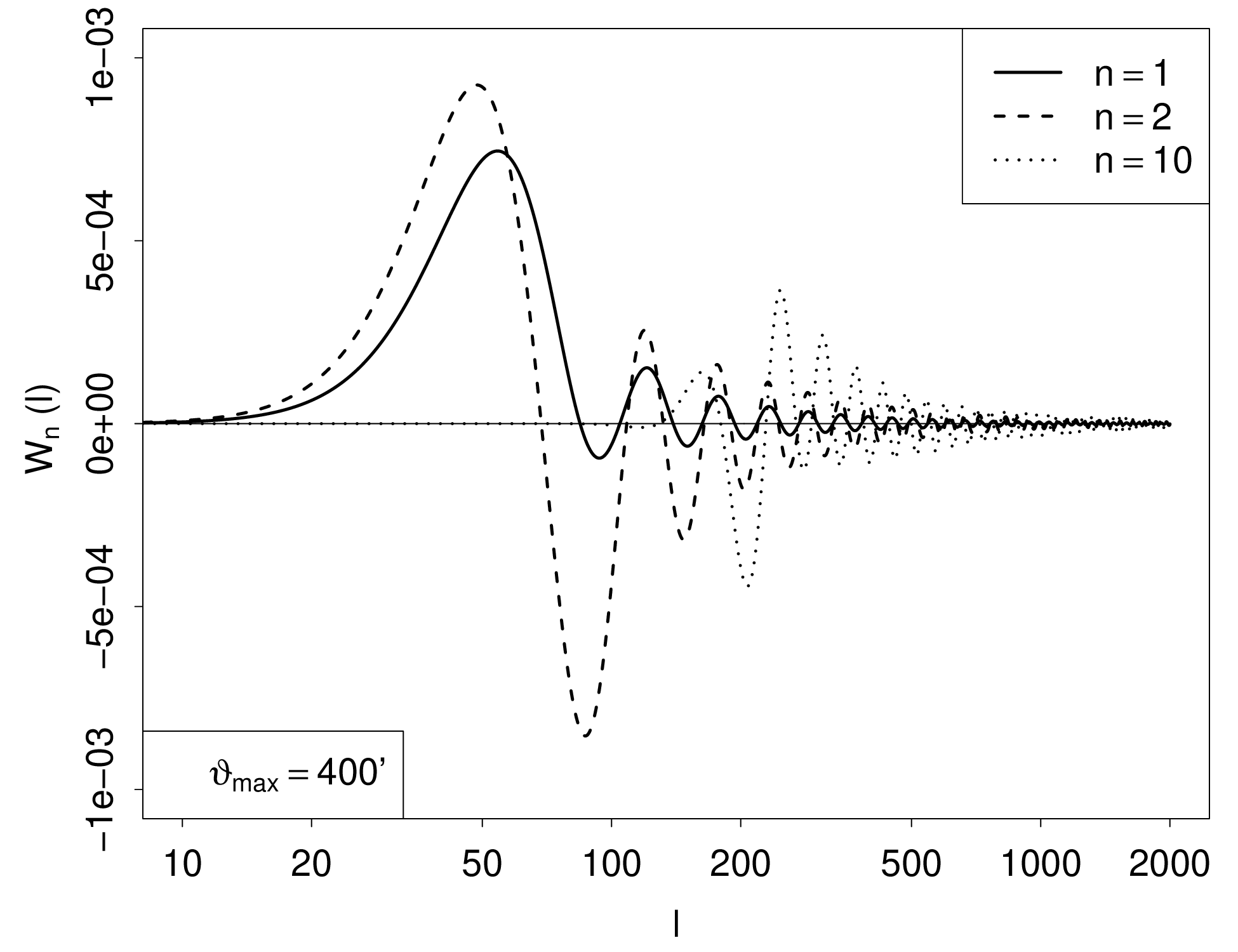}
\includegraphics[width=9cm]{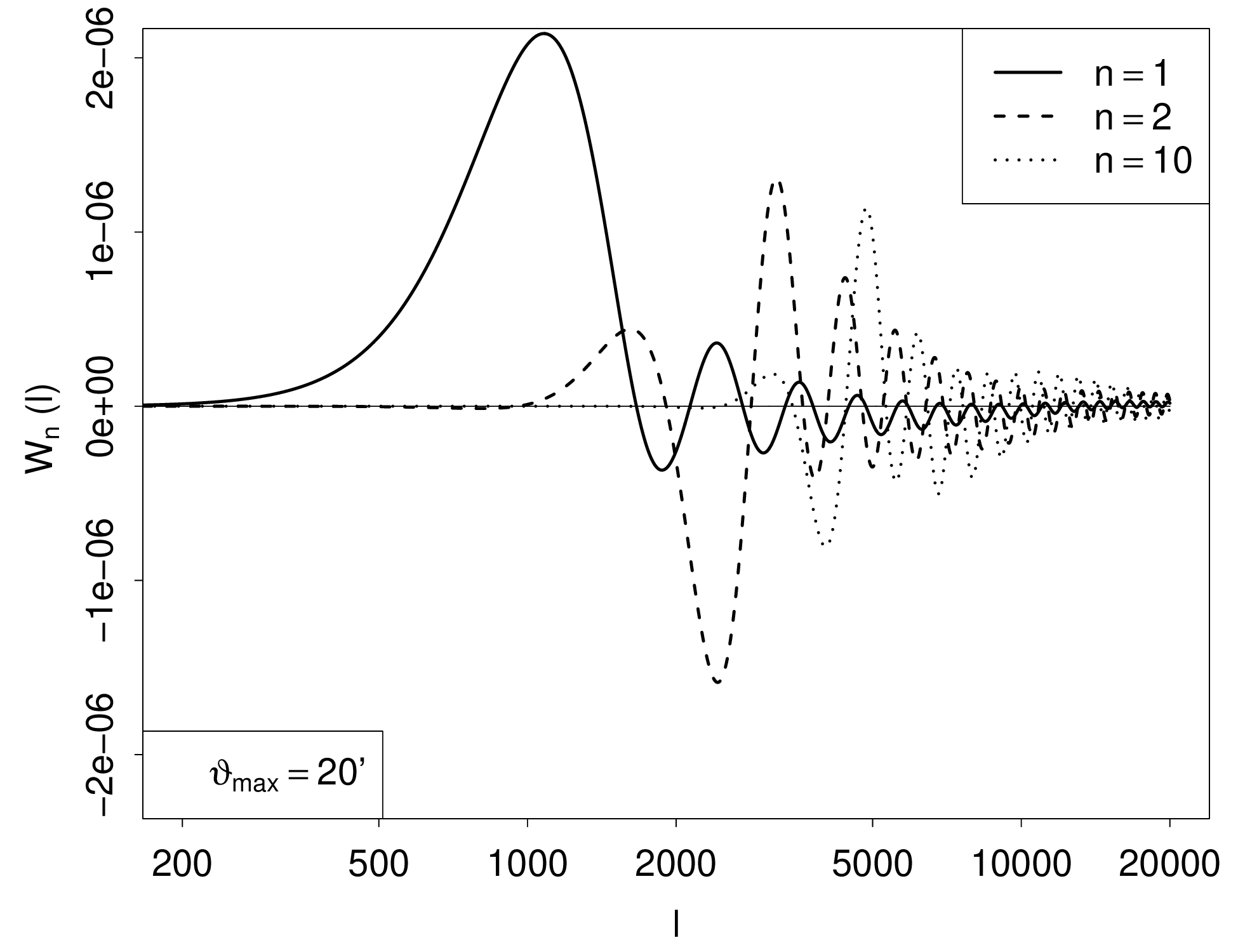} 
\caption{The functions $W_n$ as defined in Eq.\ts(\ref{eq:WnFilter})
  which relate the 
  COSEBIs to the underlying power spectrum, 
calculated from the
  $T_{\pm n}$. The 
  \textit{upper panel} corresponds to $\vartheta_\mr{max}=400'$,
  whereas the \textit{lower panel} is calculated using
  $\vartheta_\mr{max}=20'$, both for $\vt_{\rm min}=1'$}
\label{fig:W_lin}
\end{figure}

First, we construct a complete set of weight functions which are 
polynomials in $\vt$. To do so, we transform the interval $\tmin\le \vt\le \tmax$ onto the unit interval $-1\le
x\le 1$, by defining 
\be
x={2(\vt-\bar\vt)\over\Delta\vt}\;,
\elabel{PS4}
\ee
with $\bar\vt=(\tmin+\tmax)/2$, $\Delta\vt=\tmax-\tmin$. In addition,
we define the relative interval width $B=\Delta\vt/(2 \bar\vt)=
(\tmax-\tmin)/(\tmax+\tmin)$. 
Thus, as
$\vt$ varies from $\tmin$ to $\tmax$, $x$ goes from $-1$ to $+1$. Then
we set $T_{+n}(\vt)=t_{+n}(x)$, and $T_{-n}(\vt)=t_{-n}(x)$. The
$t_{+n}$ are chosen to be polynomials in $x$; as Eq. (\ref{eq:PS4}) is a
linear transformation, the polynomial order is preserved. Furthermore,
we require that the set of functions are orthonormal, i.e.,
\be
\int_{-1}^1 \d x \; t_{+n}(x)\, t_{+m}(x) = \delta_{mn} \;.
\elabel{ON-cond}
\ee
The first two functions of the set are constructed `by hand': The
lowest-order polynomial which can satisfy the constraints
(\ref{eq:conditions}) and the normalization constraint (\ref{eq:ON-cond})
is of second order. Hence, we choose $t_{+1}(x)$ to be a second-order
polynomial, and determine its three coefficients from the three
constraints. The lowest-order polynomial which can satisfy the two
constraints (\ref{eq:conditions}) and the orthonormality relation
(\ref{eq:ON-cond}) for $m=1,2$ is of third order, and its four coefficients
are determined accordingly; this yields
\bea
t_{+1}(x)\!\!&=&\!\!{1\over \sqrt{X_1}}\eck{3 B^2-5-6Bx+3(5-B^2)x^2}\;,\nonumber \\
t_{+2}(x)\!\!&=&\!\!{1\over \sqrt{X_2}}\big[B^3(25+3 B^2)-15(35+9 B^2+8 B^4)x \nonumber\\
&-&\!\!
15 B^3(3 +B^2)x^2+35(25+5 B^2+6 B^4) x^3 \big]\;,
\eea
with
\bea
X_1\!\!&=&\!\! 8(25+5 B^2+6 B^4)/5 \;,\nonumber \\
X_2\!\!&=&\!\! 8(25+5B^2+6B^4)(175+35B^2+45 B^4+B^6) \;.
\eea
To obtain the higher-order functions of this set, we note that the
Legendre polynomials $P_n(x)$ are orthogonal, and that
\[
\int_{-1}^1 \d x\;P_n(x)\,x^m =0\quad {\rm for}\;\; m<n\;.
\]
This shows that the constraints (\ref{eq:conditions}), written in terms of $x$,
are satisfied if we choose $t_+(x)\propto P_n(x)$ for all $n\ge
4$. Furthermore, the $P_n(x)$ for $n\ge 4$ are orthogonal to
$t_{+1}(x)$ and $t_{+2}(x)$, since the latter are polynomials of order
$\le 3$. Thus, choosing the normalization such as to satisfy
Eq.\ts (\ref{eq:ON-cond}), we find for $n\ge 3$,
\be
t_{+n}(x)=\sqrt{2 n+3\over 2}P_{n+1}(x)\equiv p_{n+1}(x) \;.
\elabel{tplusn}
\ee
In the upper panel of Fig.\ts\ref{fig:T_lin}, we have plotted the
filter function $T_{+n}(\vt)$ for three values of $n$. For $n\ge 3$,
they are simply proportional to the Legendre polynomials. Note that
$T_{+n}(\vt)$ has $(n+1)$ roots in the interval $\tmin\le \vt\le \tmax$,
and the normalization is chosen such that $T_{+n}(\tmax)>0$. The
corresponding filter functions $W_n(\ell)$ which relate the COSEBIs to
the power spectrum $P_{\rm E}(\ell)$ are displayed in
Fig.\ts\ref{fig:W_lin}, for several values of $n$ and for two
different values of the relative width parameter $B$ (corresponding to
two different values of $\tmax$).

For this set of functions $t_{+n}(x)$, we can obtain the corresponding
$t_{-n}(x)$ using Eq.\ts (\ref{eq:Tplusminus}),
\be
t_-(x)=t_+(x)+{4 B\over (1+B x)^2} \int_{-1}^x
\d y\;t_+(y) \,G(y,x)\;,
\elabel{tminusx}
\ee
where
\be
G(y,x)=1+ B y -3{(1+B y)^3\over (1+B x)^2} =\sum_{k=0}^3 A_k y^k\;,
\elabel{Gyx}
\ee
and the coefficients $A_k$ are given explicitly as
\bea
A_0\!&=&\! 1-{3\over (1+B x)^2}\;,\quad
A_1=B-{9 B\over (1+B x)^2}\;, \nonumber \\
A_2\!&=&\! {-9 B^2\over (1+B x)^2}\;,\quad
A_3={-3 B^3\over (1+B x)^2}\;.
\eea
For the first two functions, the integral is carried out explicitly,
yielding
\bea
t_{-1}(x)&=& {1\over\sqrt{X_1}(1+B x)^4}\sum_{k=0}^5 U_{1k} x^k\;, 
\nonumber \\
t_{-2}(x)&=& {1\over\sqrt{X_2}(1+B x)^4}\sum_{k=0}^7 U_{2k} x^k \;,
\eea
with the coefficients $U$ 
\bea
U_{10}&=&-5+19 B^2-15 B^4+3 B^6 \;,\nonumber \\
U_{11}&=& 2B(7+B^2-3 B^4) \;, \nonumber \\
U_{12}&=& 15+7 B^2+B^4-3 B^6 \;, \nonumber \\
U_{13}&=& 20 B\;, \quad
U_{14}= 10 B^2\;, \quad
U_{15}= 2 B^3\;; \nonumber
\eea
\bea
U_{20}&=& -B (350-360B^2+182 B^4-93 B^6 + 21 B^8)\;, \nonumber \\
U_{21}&=& -525 + 215 B^2-30 B^4+38 B^6+18B^8 \;, \nonumber \\
U_{22}&=& B^3(130+30 B^2 + 19 B^4 + 9 B^6) \;, \nonumber \\
U_{23}&=& 5(175 + 105 B^2 + 48 B^4 +12 B^6)   \;, \nonumber \\
U_{24}&=& 5 B (350 + 105 B^2 +87 B^4 +6 B^6) \;, \nonumber \\
U_{25}&=& B^2 (1400 +315 B^2 +339 B^4 + 6 B^6)\;, \nonumber \\
U_{26}&=& 21 B^3 (25 + 5B^2+6 B^4)\;, \nonumber \\
U_{27}&=& 3 B^4 (25 + 5 B^2 + 6 B^4) \;. \nonumber
\eea
For $n\ge 3$, we first define
\be
I_n^k(x):=\int_{-1}^x\d y\;P_n(y)\,y^k\;.
\ee
For $k=0$, one obtains
\be
I_n^0(x)={P_{n+1}(x)-P_{n-1}(x)\over 1+2n} \;,
\ee
whereas for $k\ge 1$, we make use of the recurrence relation for
Legendre polynomials, $(2n+1) y P_n(y)=(n+1)P_{n+1}(y)+n P_{n-1}(y)$,
to find
\be
I_n^k(x)={ (n+1) I_{n+1}^{k-1}(x)+ n I_{n-1}^{k-1}(x)
\over 2 n+1}\;.
\ee
Making use of Eqs.\ts (\ref{eq:tminusx}) and (\ref{eq:Gyx}), we then find, for
$n\ge 3$, 
\be
t_{-n}(x)=t_{+n}(x)
+\sqrt{2 n+3\over 2}{4 B\over (1+B x)^2}
\sum_{k=0}^3 A_k I_{n+1}^k (x) \;.
\ee
For three different values of $n$ and $\tmin=1'$, $\tmax=400'$, the
functions $T_{-n}(\vt)$ are displayed in the lower panel of
Fig.\ts\ref{fig:T_lin}. 

\subsection{Logarithmic weight functions}
Choosing the $T_{+n}$ to be polynomials in $\vt$ implies that the
structure of these weight functions is similar on all angular scales
from $\tmin$ to $\tmax$. For example, the roots of the $T_{+n}$ are
fairly evenly spread on the interval $\tmin\le \vt\le\tmax$. On the
other hand, we expect the correlation function $\xi_+(\vt)$ to show
more structure on small scales than on large scales. Hence, for a
given maximum number $N$ of modes, the large angular scales will
be sampled on finer scales than needed, whereas small angular scales
may not be sufficiently well resolved to extract all information
contained in the correlation function.

\begin{figure*}
\begin{verbatim}
Nmax=20; tmin=1; tmax=400; zm=Log[Rationalize[tmax/tmin]]
gamm[a_,z_]=Gamma[a,0,z]
Do[J[k,j]=Re[N[gamm[j+1,-k zm]/(-k)^(j+1),130]],{k,1,2},{j,0,2 Nmax+1}]
Do[J[4,j]=Re[N[gamm[j+1,-4 zm]/(-4)^(j+1),130]],{j,0,2 Nmax+1}]
Do[
  Do[a[n,j]=J[2,j]/J[2,n+1]; a[n+1,j]=J[4,j]/J[4,n+1],{j,0,n}]; b[n]=-1; b[n+1]=-1;
  Do[a[m,j]=NSum[J[1,i+j] c[m,i],{i,0,m+1}, WorkingPrecision->80, NSumTerms->Nmax],{m,1,n-1},{j,0,n}];
  Do[bb[m]=-NSum[J[1,i+n+1] c[m,i],{i,0,m+1}, WorkingPrecision->80, NSumTerms->Nmax],{m,1,n-1}];
  Do[a[m,j]=a[m,j]/bb[m],{m,1,n-1},{j,0,n}]; Do[b[m]=1,{m,1,n-1}];
  A=Table[a[i,j],{i,1,n+1},{j,0,n}]; B=Table[b[i],{i,1,n+1}];
  CC=LinearSolve[A,B]; Do[c[n,j]=CC[[j+1]],{j,0,n}]; c[n,n+1]=1;
  tt[n,z_]=Simplify[Sum[c[n,j] z^j,{j,0,n+1}]];
  roots=NSolve[tt[n,z]==0,z];Do[r[n,j]=N[roots[[j,1,2]],8],{j,1,n+1}];
  t[n,z_]=Product[(z-r[n,j]),{j,1,n+1}];
  normgral=NIntegrate[Exp[z] t[n,z]^2,{z,0,zm},WorkingPrecision->50];
  norm[n]=Sqrt[(Exp[zm]-1)/normgral]; t[n,z_]=t[n,z] norm[n],
{n,1,Nmax}]
ROOTS=Table[r[n,j],{n,1,Nmax},{j,1,Nmax+1}]
\end{verbatim}
\caption{Mathematica \citep{WolfMathe} program to calculate the roots in
  Eq.\ts(\ref{eq:tpluslogprod}) -- they are stored with 8 significant
  digits in the lower left halve of the table {\tt ROOTS}. Furthermore,
the array {\tt norm[n]} contains the normalization coefficients $N_n$}
\flabel{matheprog}
\end{figure*}

\begin{figure*}
\sidecaption
\includegraphics[width=13cm]{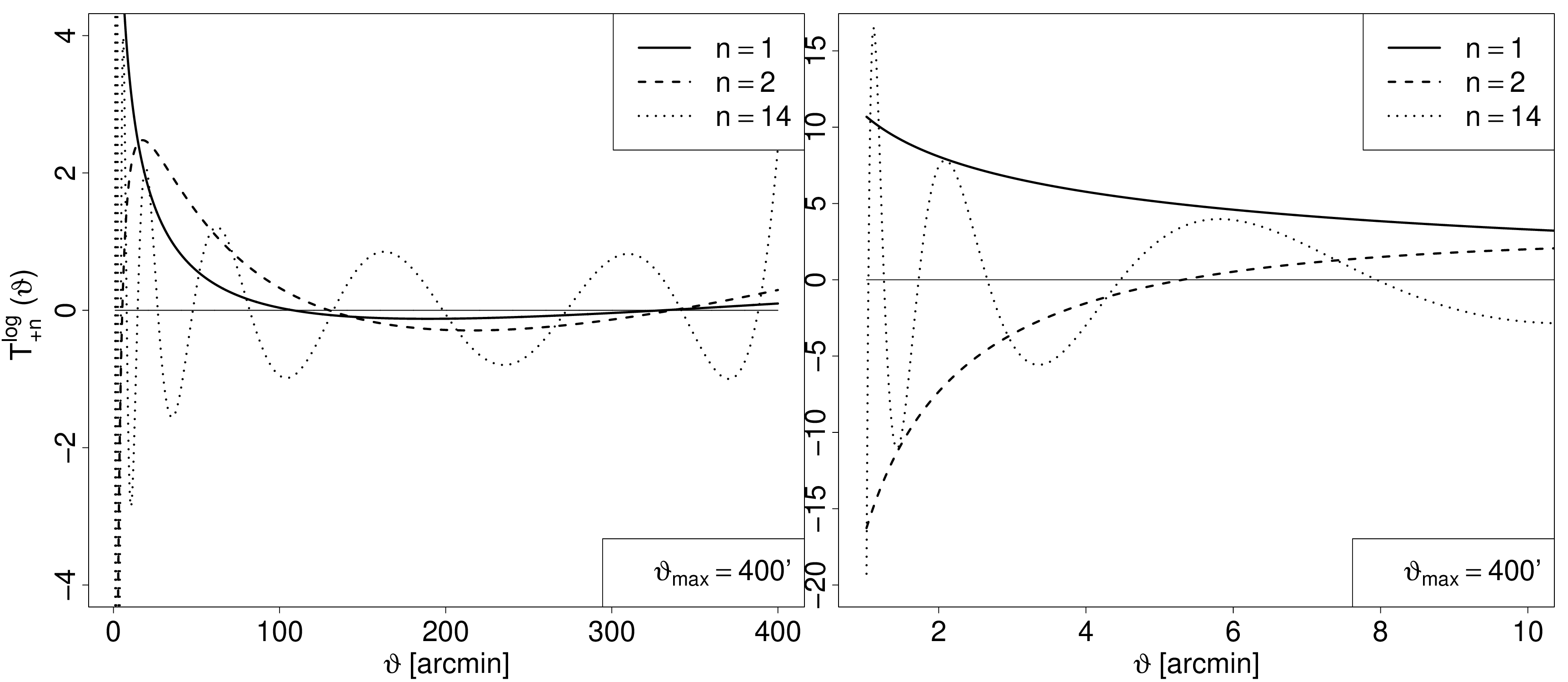}
  \caption{The logarithmic filter functions $T_{+n}^\mr{log}$ for
    $\vt_{\rm min}=1'$ and $\vartheta_\mr{max}=400'$. The left panel
    shows the function over the whole interval, whereas the right
    panel provides a more detailed view for small $\vt$}
         \label{fig:T_log_plus}
\end{figure*}

\begin{figure*}
\sidecaption
\includegraphics[width=13cm]{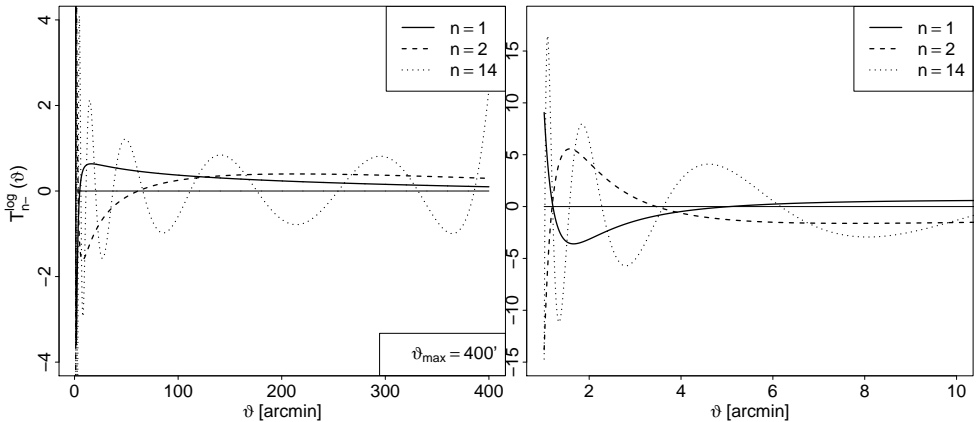}
  \caption{The logarithmic filter functions $T_{-n}^\mr{log}$ for
    $\vt_{\rm min}=1'$ and $\vartheta_\mr{max}=400'$. As in
    Fig.\ts\ref{fig:T_log_plus}, the left panel
    shows the function over the whole interval, whereas the right
    panel provides a more detailed view for small $\vt$}
         \label{fig:T_log_minus}
\end{figure*}

In order obtain a finer sampling of the small-scale correlation
function for a given $N$, we now construct a set of weight functions
which are polynomials in $\ln\vt$.  Hence the roots of these weight
functions are approximately evenly spaced in $\ln\vt$, thus the weight
functions sample small angular scales with higher resolution than
large angular scales.  As in Sect.\ts\ref{sc:polynomials}, this set
of functions must fulfill the constraints (\ref{eq:conditions}), and
we require the functions to be orthonormal. Hence, the lowest-order
weight function again is of second-order. We parametrize this set of
weight functions as
\be
t_{+n}^\mr{log}(z) = \sum_{j=0}^{n+1} c_{nj}z^j
=N_n \sum_{j=0}^{n+1} \bar c_{nj} z^j \;,
\elabel{Tlog}
\ee
where we choose
\be 
z = \ln \left(\vt/\tmin\right)\;,
\ee
which varies from $ 0 $ to $z_{\mr{max}}= \ln(\tmax/\tmin)$ as $\vt$
goes from $\tmin$ to $\tmax$. Furthermore, we defined $c_{nj}=N_n \bar
c_{nj}$ with $N_n\equiv c_{n(n+1)}\ne 0$, so that $\bar
c_{n(n+1)}=1$. In this way, the relative amplitude of the $c$'s is
decoupled from the overall normalization $N_n$. 
As before, we set $T_{+n}^\mr{log}(\vt)
= t_{+n}^\mr{log}(z)$ and $T_{-n}^\mr{log}(\vt) = t_{-n}^\mr{log}(z)$.
With this transformation of variables the constraints
(\ref{eq:conditions}) become 
\be
 \int_0^{z_{\mr{max}}} \d z\;  
\mr e^{2z}\,t_{+n}^\mr{log}\left(z\right) = 0 
= \int_0^{z_{\mr{max}}} \d z \; 
\mr e^{4z}\,t_{+n}^\mr{log}\left(z\right) \;,
\elabel{conditionslog}
\ee
and an orthonormality condition analogous to Eq.\ts (\ref{eq:ON-cond})
can be written as
\bea
\frac{1}{\Delta \vt}\!\!\!\!\!&&\!\!\!\!\!  \int_\tmin^\tmax \d \vt \; 
T_{+n}^\mr{log}(\vt) T_{+m}^\mr{log}(\vt) \nonumber \\
&=&\!\! \frac{\tmin}{\Delta \vt}\int_0^{z_{\mr{max}}} \d z\; 
\mr{e}^z t^\mr{log}_{+n}(z) t^\mr{log}_{+m}(z) 
= \delta_{nm}\;.
\elabel{ON-log}
\eea
To write these constraints in a more compact form we 
define the set of coefficients
\be
J(k,j)=\int_0^{z_{\mr{max}}}\d z\; \mr{e}^{k z}\,z^j
={ \gamma(j+1,-k z_{\mr{max}})\over (-k)^{j+1}}\;,
\ee
where $\gamma(a,x)$ is the incomplete Gamma function. 


With the representation (\ref{eq:Tlog}), the constraints
(\ref{eq:conditionslog}) become
\bea
\sum_{j=0}^n \bar c_{nj}\,J(2,j)&=& -J(2,n+1) \;, \nonumber \\
\sum_{j=0}^n \bar c_{nj}\,J(4,j)&=& -J(4,n+1) \;.
\elabel{const2+4}
\eea
These two equations determine the two coefficients $\bar c_{10}$, $\bar
c_{11}$ needed to obtain $t^\mr{log}_{+1}(z)$. We then obtain the
corresponding coefficients $\bar c_{nj}$ by iterating in $n$. Thus,
for a given $n$, we assume that the $\bar c_{mj}$ have been determined
for all $m<n$. Then, the $\bar c_{nj}$ are obtained from the two
Eqs.\ts (\ref{eq:const2+4}), and the $(n-1)$ orthogonality conditions
(\ref{eq:ON-log}) for $1\le m\le n-1$, which read in the
representation (\ref{eq:Tlog})
\[
\sum_{j=0}^{n+1} \sum_{i=0}^{m+1}  J(1,i+j)\,\bar c_{mi} \,
\bar c_{nj} =0 \;,
\]
or
\be
\sum_{j=0}^n \rund{\sum_{i=0}^{m+1} J(1,i+j)\,\bar c_{mi} }
\bar c_{nj} = -
\sum_{i=0}^{m+1} J(1,i+n+1)\,\bar c_{mi} \;,
\elabel{MatEqforc}
\ee
where we used that $\bar c_{n(n+1)}=1$.  Thus, together we have $n+1$
linear equations for the $n+1$ unknown coefficients $\bar c_{nj}$,
$0\le j\le n$, which in principle can be readily solved (but see
below). Finally, to obtain the normalization of the functions, we use
Eq.\ts (\ref{eq:ON-log}) for $m=n$, which together with Eq.\ts
(\ref{eq:Tlog}) yields
\be
N_n^2 \sum_{i,j=0}^{n+1} \bar c_{ni}\,\bar c_{nj}\,J(1,i+j)
={\Delta\vt\over \vt_{\rm min}}={\rm e}^{z_{\rm max}}-1\;,
\ee
which determines $N_n$ (and thus the
$c_{nj}=N_n \bar c_{nj}$) up to an (arbitrary) sign. For definiteness,
we choose the sign such that $t^\mr{log}_{+n}(z_{\rm max})>0$,
implying that $N_n=c_{n(n+1)}>0$.   

It turns out that the solution of the system of linear equations for
the $c$'s requires very high numerical accuracy for even moderately
large $n$, in particular for large values of $\tmax/\tmin$. We used
Mathematica \citep{WolfMathe} with large setting of {\tt
  WorkingPrecision} for calculating the incomplete Gamma function and
for carrying out the sums in Eq. (\ref{eq:MatEqforc}). Once the $c$'s have
been determined, the integrals in Eqs.\ts (\ref{eq:conditionslog}) and
(\ref{eq:ON-log}) -- the latter for $m<n$ --  have been calculated to 
check the accuracy of the solution. We found that, for
$\tmax/\tmin=400$ and for $n_{\rm max}=20$, one needs to determine the
$c$'s to 40 significant digits, in order for all these integrals, which
should be zero, to attain values less than $0.1$. We then calculated
the $n+1$ roots $r_{n,i}$ of the $t_{+n}^\mr{log}(z)$, and represented
the functions as
\be
 t_{+n}^\mr{log}(z) =N_n \prod_{i=1}^{n+1} (z-r_{ni})\;.
\elabel{tpluslogprod}
\ee
For the same parameters as before,
using only five significant digits for the $r$'s renders all the
integrals zero to better than $10^{-6}$, and with eight significant
digits, the integrals are zero to better than $10^{-17}$ even for
$n_{\rm max}=40$. Thus, the representation (\ref{eq:tpluslogprod}) is
the adequate one for practical work. A short Mathematica program for
calculating the $r_n$ is displayed in Fig.\ts\ref{fig:matheprog}.

\begin{figure}
\includegraphics[width=9cm]{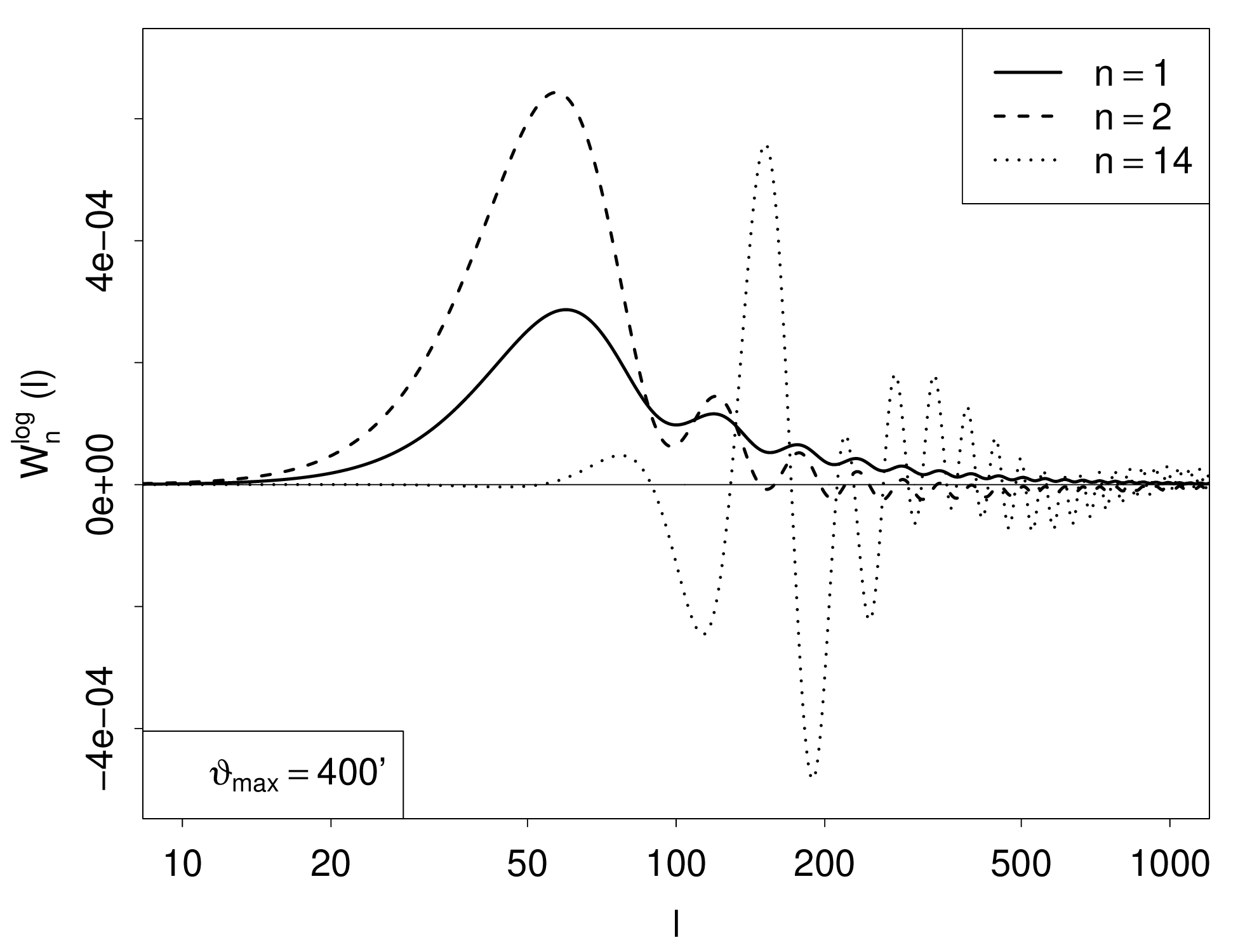}
\includegraphics[width=9cm]{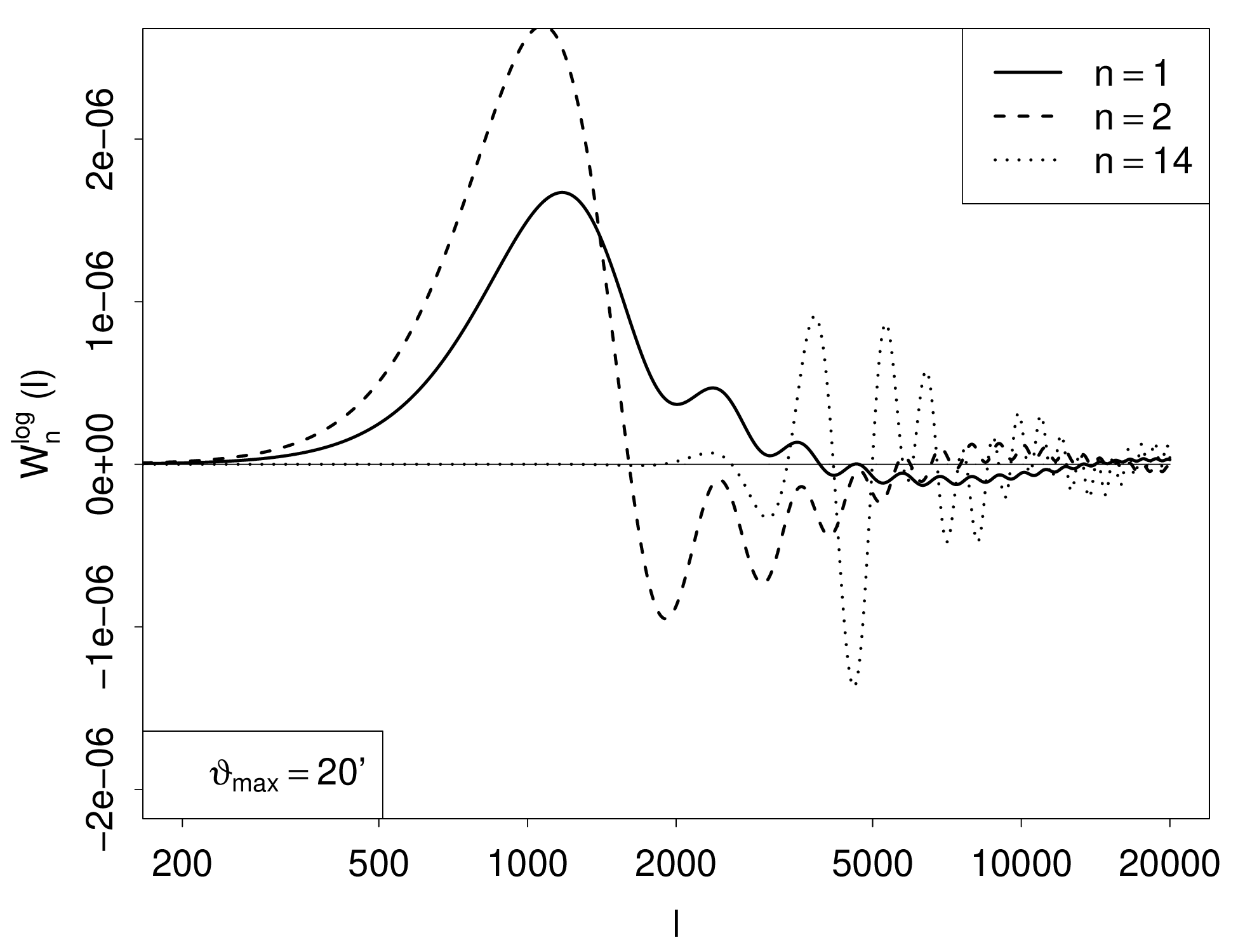}
\caption{The $W_n$-functions calculated from $T_n^\mr{log}$. The
  \textit{upper panel} corresponds to $\vartheta_\mr{max}=400'$,
  whereas the \textit{lower panel} is calculated using
  $\vartheta_\mr{max}=20'$, and $\vt_{\rm min}=1'$ in both cases}
\label{fig:Wlog}
\end{figure}

The corresponding $T_{-n}^\mr{log}$ are constructed from
Eq.\ts (\ref{eq:Tplusminus}), by defining $y=\ln(\theta/\vt_{\rm
  min})$, which yields
\bea
t_{-n}^\mr{log}(z) &=& t_{+n}^\mr{log}(z) 
+4\int_0^z\d y\;t_{+n}^\mr{log}(y)  
\rund{ {\rm e}^{2(y-z)} - 3 {\rm e}^{4(y-z)} }\nonumber\\
&=& t_{+n}^\mr{log}(z) 
  4 \sum_{j=0}^{n+1} c_{nj} \int_0^z\d y\; y^j
\rund{ {\rm e}^{2(y-z)} - 3 {\rm e}^{4(y-z)} } \nonumber\\
&=& 
t_{+n}^\mr{log}(z) 
 4 {\rm e}^{-2 z} \sum_{j=0}^{n+1} { c_{nj} \over (-2)^{j+1}} 
\Bigl[ \gamma(j+1,-2 z) \\
&&-  { 3 {\rm e}^{-2 z}\over 2^{j+1}}\,\gamma(j+1,-4 z) \Bigr]\;.\nonumber
\eea
Given the remarks above, the first of these expressions (i.e.,
numerical integration) is the method of choice if the
$t_{+n}^\mr{log}(z)$ are given in the form 
(\ref{eq:tpluslogprod}). Alternatively, making use of the
representation
\[
\gamma(j+1,z)=j!\eck{1-{\rm e}^{-z}\sum_{m=0}^j {z^m\over m!}} \;,
\]
one can write the $t_{-n}^\mr{log}(z)$ as
\be
t_{-n}^\mr{log}(z)=a_{n2}{\rm e}^{-2z}-
a_{n4}{\rm e}^{-4z}+\sum_{m=0}^n d_{nm} z^m\;,
\ee
where the coefficients are given as
\bea
a_{n2}\!\! &=& \!\! 4 \sum_{j=0}^{n+1}{c_{nj}\,j!\over (-2)^{j+1}}\;,\quad
a_{n4}=12 \sum_{j=0}^{n+1}{c_{nj}\,j!\over (-4)^{j+1}}\;, \nonumber\\
d_{nm}\!\! &=&\!\! c_{nm}+{4\over m!}\sum_{j=m}^{n+1} c_{nj}\,j! (-2)^{m-j-1}
\rund{3\; 2^{m-j-1} -1} \;.  
\eea
In Figs.\ts\ref{fig:T_log_plus} and \ref{fig:T_log_minus}, we have
plotted the filter functions $T^{\rm log}_{\pm n}$ for $\tmin=1'$ and
$\tmax=400'$. The left panels show these filter functions over the
whole angular range, the right panels show an enlargement for small
values of $\vt$. As expected, the roots of the weight functions are
clustered towards lower values of $\vt$. Thus, for a fixed maximum
number of $n$, these functions resolve those scales better than the
linear filter functions. Figure \ts\ref{fig:Wlog} shows the filter
functions $W_n(\ell)$ which, according to Eq.\ts(\ref{eq:EBfromP}),
relates the COSEBIs to the underlying power spectrum $P_{\rm
  E}(\ell)$. With increasing $n$, the COSEBIs are sensitive to power
at increasingly larger values of $\ell$.

\begin{figure}
 \includegraphics[width=9cm]{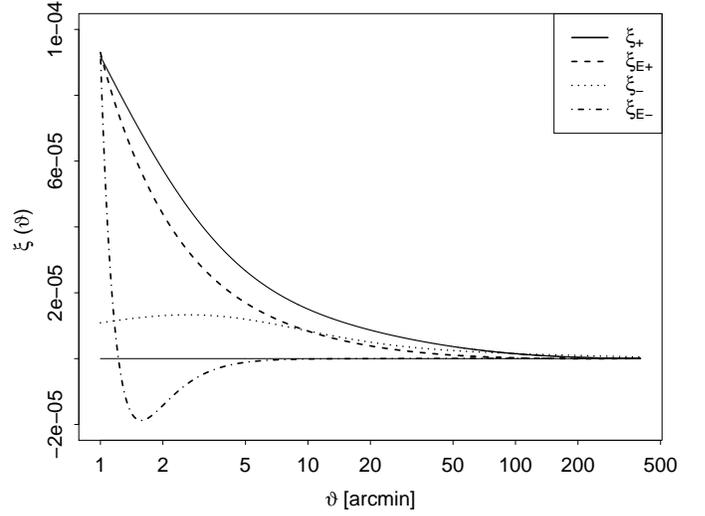}
\caption{The 2PCFs $\xi_\pm(\vt)$ and the corresponding pure E-mode
  correlation functions $\xi^{\rm E}_\pm(\vt)$, for $\tmin=1'$,
  $\tmax=400'$, and the fiducial cosmological model described in
  Sect.\ts\ref{sc:likelihoodanalysis}}
\flabel{xiEpm}
\end{figure}

\subsection{E-/B-mode correlation functions}
\cite{cnp02} and \cite{svm02} constructed E-/B-mode correlation
functions, which consist of the original correlation function
$\xi_\pm(\vt)$ plus a correction term which is again an integral over
correlation functions. However, these correction terms are
unobservable, since the integral extends over an infinite angular
range. Thus, these E-/B-mode correlation functions cannot be obtained
in practice and are of little use.

With the full E-/B-mode decomposition provided by the COSEBIs, we can
define new pure E-/B-mode correlation functions, 
\bea
\xi^{\rm E}_\pm(\vt)&=&{2\over \vt\,\Delta\vt}\sum_{n=1}^\infty E_n\,T_{\pm
  n}(\vt)\;, \nonumber \\
\xi^{\rm B}_\pm(\vt)&=&{2\over \vt\,\Delta\vt}\sum_{n=1}^\infty B_n\,T_{\pm
  n}(\vt)\;; 
\elabel{EBpureFCs}
\eea
obviously, the $\xi^{\rm E}_\pm$ only depend on the E-mode shear,
whereas the $\xi^{\rm B}_\pm$ contains information only from B-modes.
Owing to the constraints (\ref{eq:conditions}) which the functions
$T_{+n}$ have to obey, one finds that 
\be
\int_{\tmin}^{\tmax}\d\vt\;\vt\,\xi^{\rm E}_+(\vt)
=0=
\int_{\tmin}^{\tmax}\d\vt\;\vt^3\,\xi^{\rm E}_+(\vt) \;.
\elabel{xiEpluconstraints}
\ee
In fact, as shown in SK07, the function $T_-$ also obeys analogous
constraints, namely
\[
\int_{\tmin}^{\tmax}{\d\vt\over \vt}\,T_-(\vt)= 0=
\int_{\tmin}^{\tmax}{\d\vt\over \vt^3}\,T_-(\vt) \;,
\]
so that
\be
\int_{\tmin}^{\tmax}{\d\vt\over \vt}\,\xi^{\rm E}_-(\vt)
=0=
\int_{\tmin}^{\tmax}{\d\vt\over \vt^3}\,\xi^{\rm E}_-(\vt) \;.
\elabel{xiEminconstraints}
\ee
In Fig.\ts\ref{fig:xiEpm}, we have plotted the pure E-mode correlation
functions $\xi^{\rm E}_\pm$, together with the orinial 2PCFs
$\xi_\pm$, for a fiducial $\Lambda$CDM cosmological model that
  will be described in the next section; the overall shape of these
  functions, however, does not depend on the details of the choice of
  cosmological parameters.  Although not easily visible, $\xi^{\rm
  E}_\pm$ both have two roots, as required by the constraints
(\ref{eq:xiEpluconstraints}) and (\ref{eq:xiEminconstraints}). The
function $\xi^{\rm E}_+$ is rather similar in shape to the original
2PCF $\xi_+$, modified in a way as to obey
Eq.\ts(\ref{eq:xiEpluconstraints}). However, $\xi^{\rm E}_-$ has a
very different shape than $\xi_-$. In fact, it is easy to see from
Eqs.\ts(\ref{eq:Tplusminus}) and (\ref{eq:conditions}) that $\xi^{\rm
  E}_-(\tmin)= \xi^{\rm E}_+(\tmin)$, $\xi^{\rm E}_-(\tmax)= \xi^{\rm
  E}_+(\tmax)$.  In Appendix \ts\ref{sc:EBmodeCFs}, we show how these
new pure-mode correlation functions are related to the original
2PCFs. As is obvious from their definition, these pure-mode
correlation functions can be obtained from the 2PCFs over a finite
interval, hence their estimation does not require extrapolations or
`inventing data'.

\begin{figure}
 \includegraphics[width=9cm]{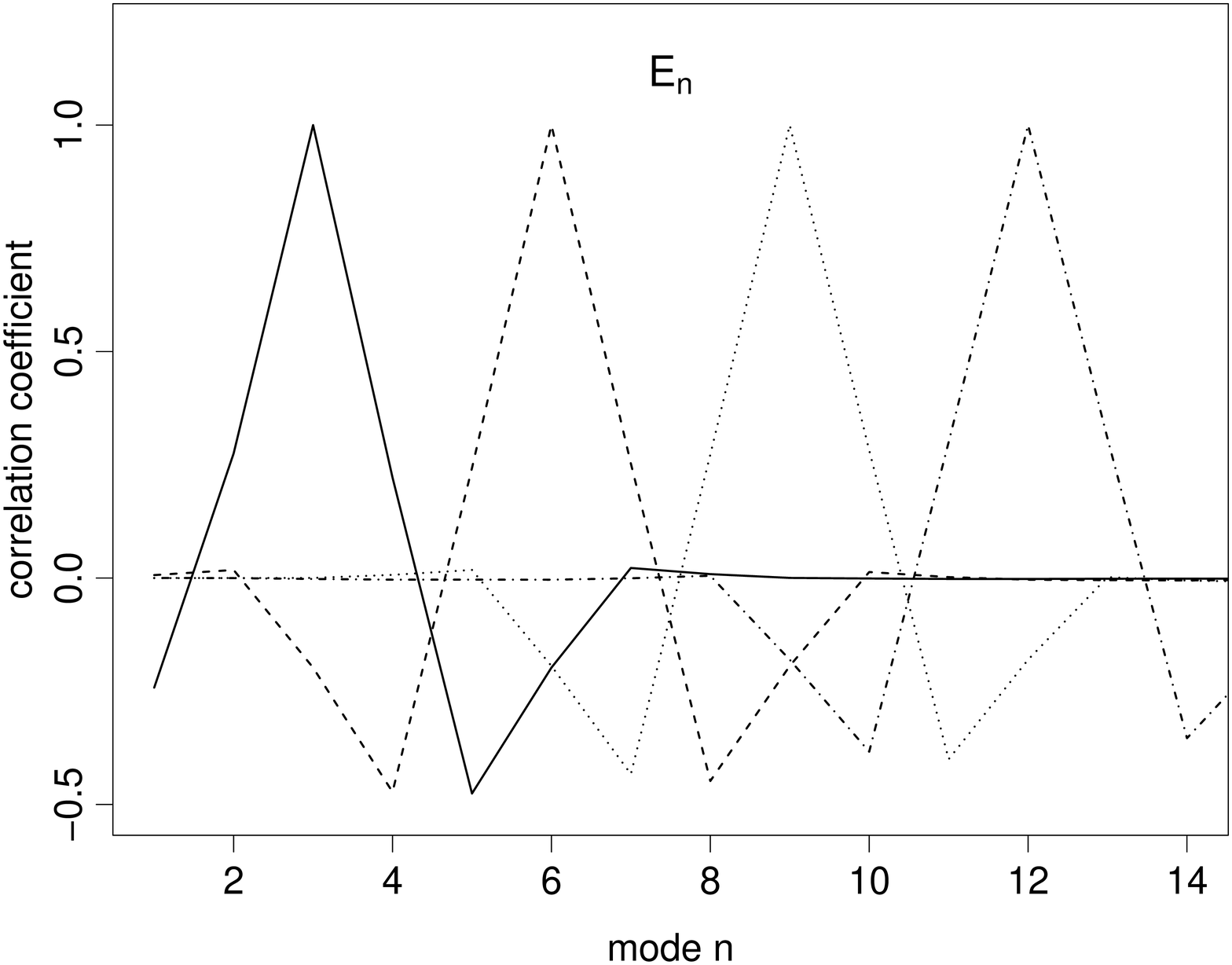}
\includegraphics[width=9cm]{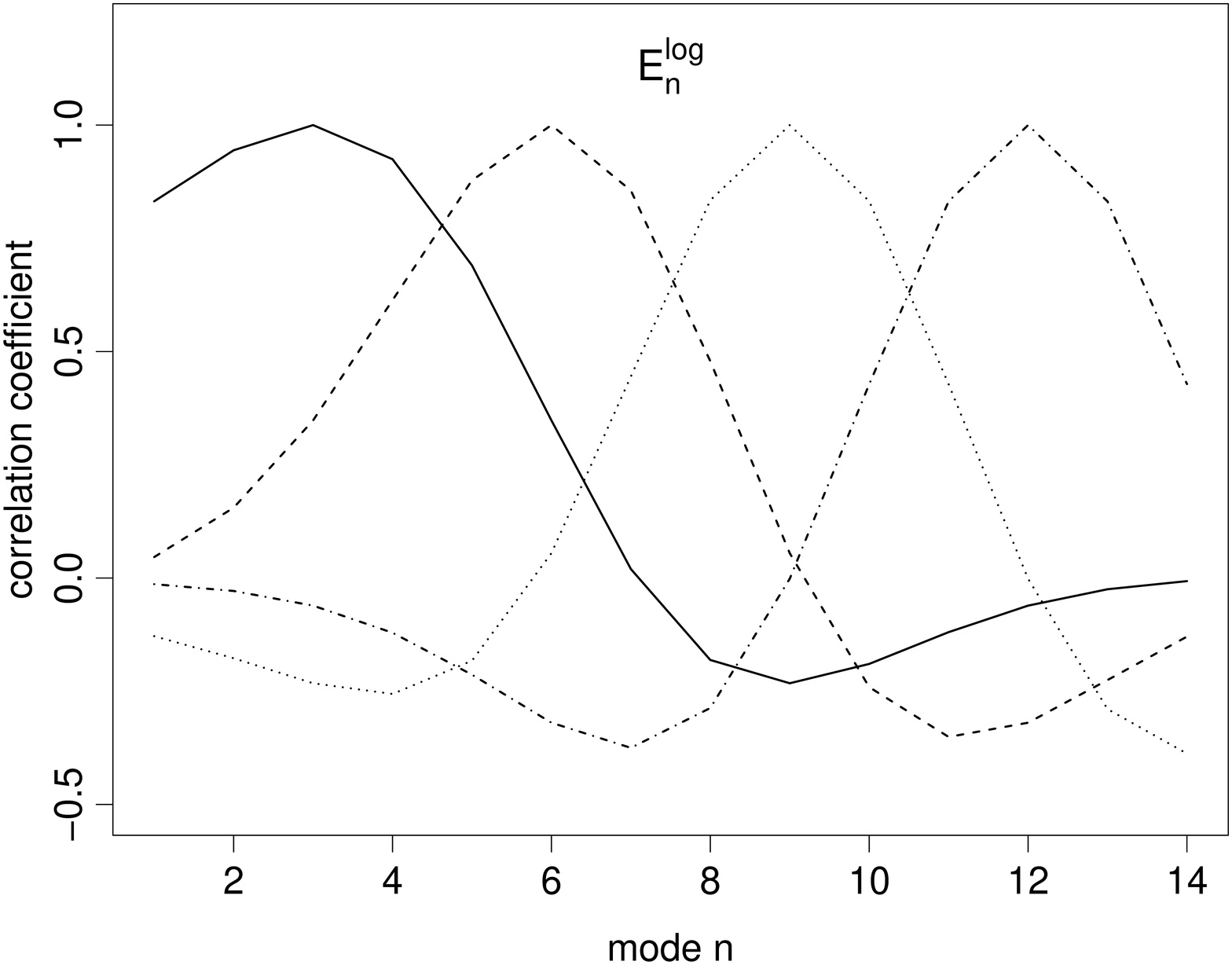}
    \caption{The correlation coefficients (\ref{eq:correlrmn})
for linear
      (\textit{top}) and logarithmic (\textit{bottom}) weight
      functions $T_{\pm n}$, calculated for $\tmin=1'$, $\tmax=400'$,
      and the fiducial cosmological model described in the text} 
         \label{fig:correlation}
\end{figure}

\begin{figure*}
\sidecaption
 \includegraphics[width=14cm]{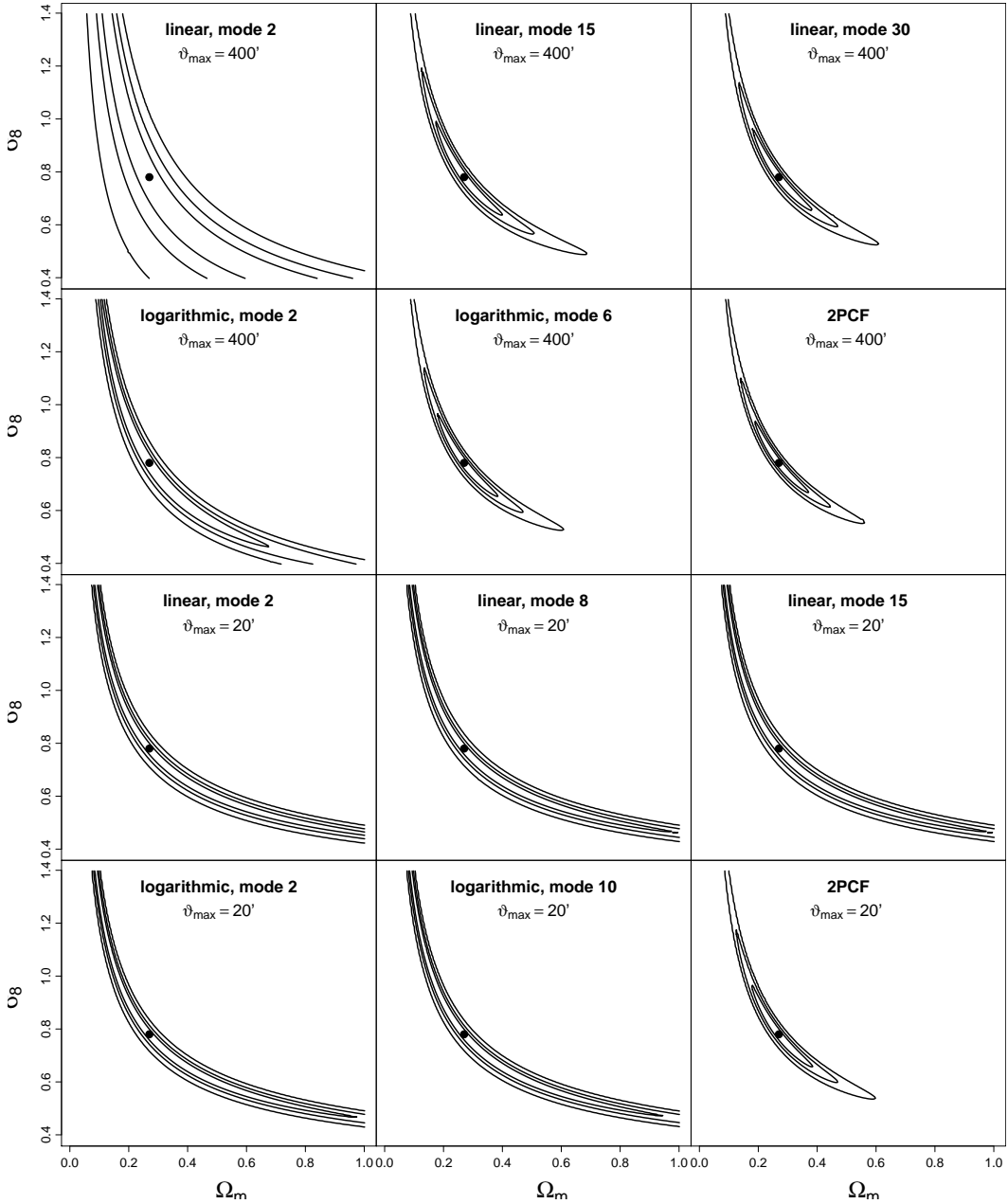}
 \caption{Likelihood contours for a fiducial cosmic shear survey, with
   parameters described in Sect.\ts\ref{sc:ModelChoice}. The upper
   (lower) six panels correspond to $\vartheta_\mathrm{max}=400'$
   ($20'$). Shown in the first and third rows are the likelihood as
   obtained from the COSEBIs with linear filter functions and various
   $n_{\rm max}$, in the second and fourth rows the likelihood as
   obtained from the logarithmic filter functions, and in comparison,
   we show the likelihood obtained from the shear two-point
   correlation functions}
        \label{fig:contour_EB}
\end{figure*}

\begin{figure*}
 \includegraphics[width=9cm]{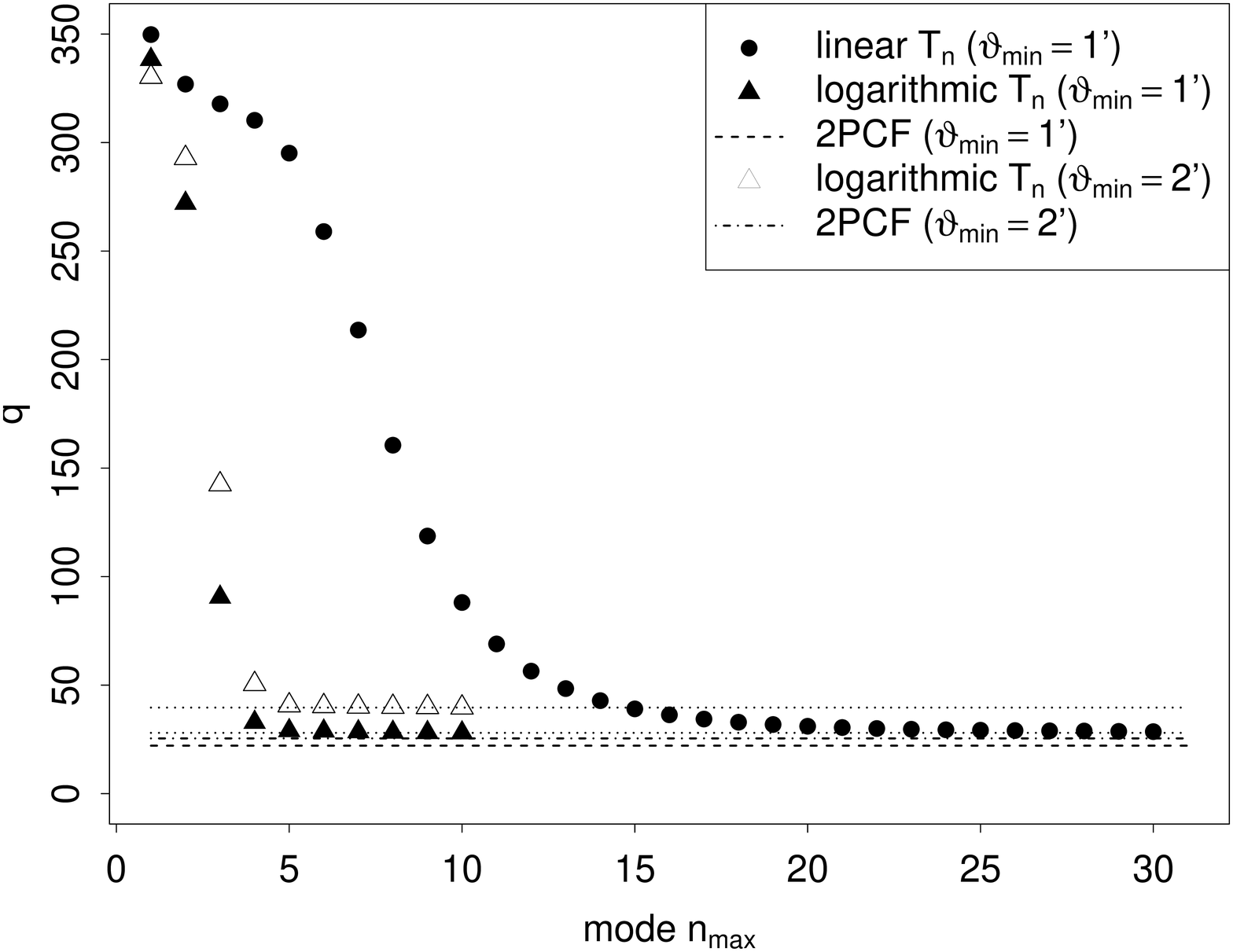}
\includegraphics[width=9cm]{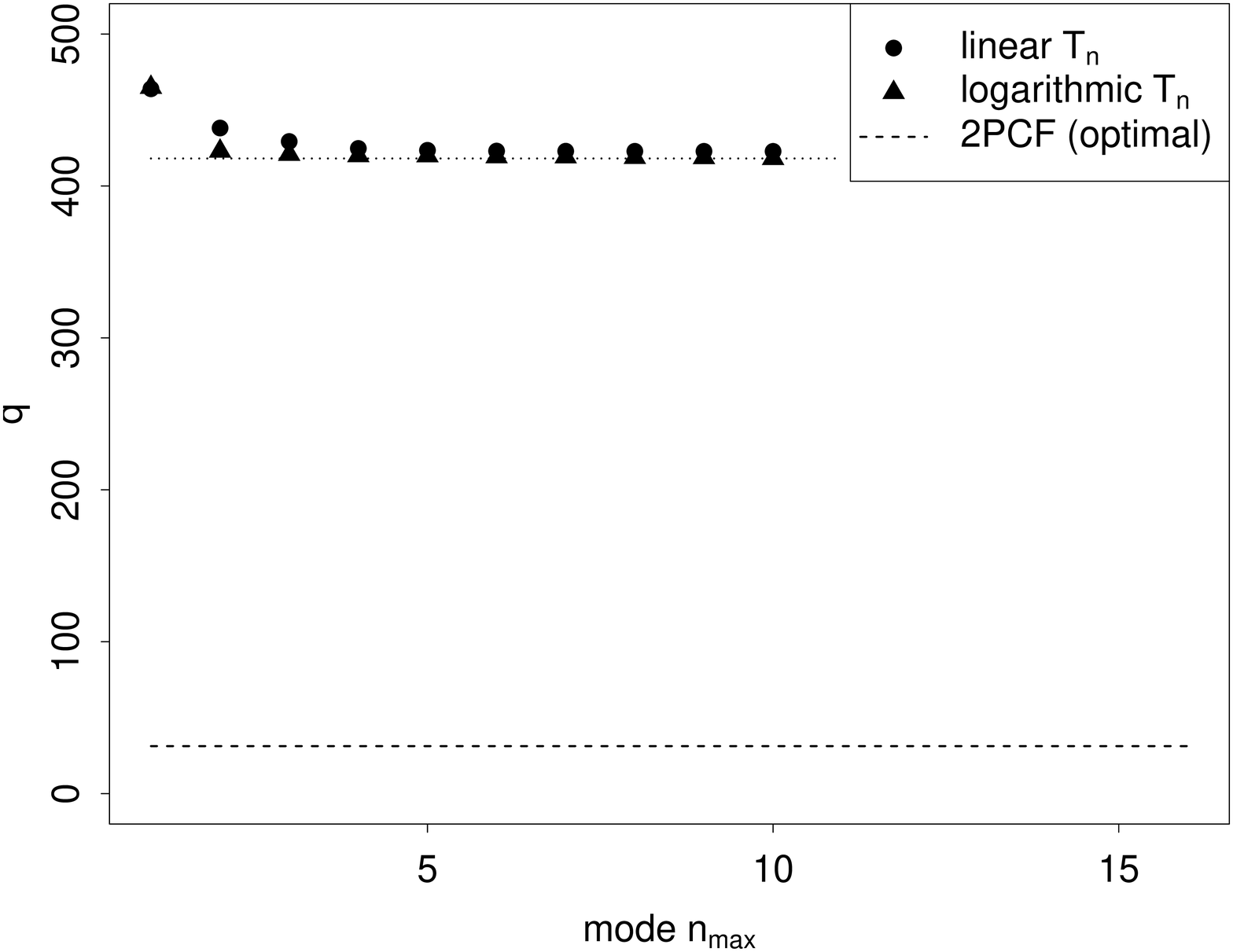}
\caption{The values of $q$ -- see Eq.\ts(\ref{eq:detquadrups}) --
  calculated from the COSEBIs for the case of linear
  (\textit{circles}) or logarithmic (\textit{triangles})
  $T_n$-functions, as a function of the maximum mode $n_{\rm max}$
  which was included in the likelihood analysis.  The results in the
  \textit{left} (\textit{right}) panel correspond to
  $\vartheta_\mr{max}=400'$ ($\vartheta_\mr{max}=20'$), and the filled
  symbols are calculated for $\tmin=1'$; in the left panel, we also
  plot corresponding results for $\tmin=2'$, indicated by the open
  triangles.  The dashed (dash-dotted) line represents the optimal $q$
  for $\tmin=1'$ ($\tmin=2'$), obtained when using the 2PCFs
  directly. The dotted lines shows the asymptotic value of $q$
  achieved for large $n_{\rm max}$  }
         \label{fig:EBq}
\end{figure*}

\section{\llabel{likelihoodanalysis} Likelihood analysis}
We calculate the posterior likelihood in the $\om$-$\sig$ parameter
space for four cases of COSEBIs ($\vt_\mr{max}=400'$,
$\vt_\mr{max}=20'$, each for $T_n^\mr{log}$ and $T_n^\mr{lin}$). Note
that, unless stated otherwise, we choose $\vt_\mr{min}=1'$ as the minimum
separation in the 2PCF. For each of the four cases we are interested
in two main questions: First, how does the information content evolve
when including more modes $n$ in the likelihood analysis? Second, once
it saturates, how large is the difference to the information content
of the 2PCFs? 

\subsection{\llabel{ModelChoice}Model choice}
In the likelihood analysis we assume a flat universe, and vary the
matter density $\om$ (and simultaneously $\Omega_\Lambda=1-\om$ to
preserve flatness) and the normalization $\sigma_8$ of the density
fluctuations; all other parameters are held fixed, i.e. the
dimensionaless Hubble constant $h=0.73$, the density parameter in
baryons $\Omega_\mr b=0.044$, and the slope of the primordial
fluctuation power spectrum $n_\mr s=1.0$. We choose $\om=0.27$ and
$\sigma_8=0.78$ as our fiducial model which enters the likelihood
analysis in this section and represents the cosmological model used in
Fig.\ts\ref{fig:xiEpm}. The B-mode power spectrum is set to zero,
$P_{\rm B}(\ell)\equiv 0$, whereas the shear power spectra $P_\mr E$
are obtained from the three-dimensional density power spectra
$P_\delta$ using Limber's equation \citep[see, e.g.,][]{Kais98}. The
power spectrum $P_\delta$ is calculated with the transfer function
from \cite{ebw92}. For the non-linear evolution we use the fitting
formula of \cite{sm03}. In the calculation of $P_\mr E$ we choose a
redshift distribution of source galaxies similar to that of
\cite{bhs07},
\be 
n(z)=\frac{\beta}{z_0 \Gamma \left( \left(1+\alpha \right)/\beta \right)} \left( \frac{z}{z_0}\right)^\alpha \exp \left[ - \left(  \frac{z}{z_0} \right)^\beta \right]\,,
\elabel{redshiftben}
\ee 
with $\alpha=0.836$, $\beta=3.425$, $z_0=1.171$. The corresponding
2PCFs are calculated from Eq.\ts(\ref{eq:xi+-}), and from these, the
COSEBIs are calculated according to Eq. (\ref{eq:EBmodes}) for various
modes $n$ using linear and logarithmic filter functions.  The
covariances used in our likelihood analysis are calculated from the
power spectrum $P_\mr E$ as described in \cite{jse08}, assuming our
fiducial cosmology. This method does not account for the
non-Gaussianity of the shear field or the cosmology-dependence of the
covariance \citep{esh09}, however these issues are not crucial for our
purpose as we are only interested in the relative performance of
COSEBIs and the 2PCFs. More important is that we can choose an
arbitrary binning in the 2PCF covariance. The latter aspect in
combination with the speed of the calculation is decisive to resolve
the numerical issues in the calculation of the COSEBIs'
covariance. The survey parameters read $A=170$ $\mr{deg}^2$,
$n_\mr{gal}=13.3/\mr{arcmin}^2$, and $\sigma_\epsilon=0.42 $, and
correspond to those of the upcoming cosmic shear analysis of the full
CFHTLS survey area.

The exact method to calculate the posterior likelihood from the data
vectors and covariances is described in \cite{esk09}. Similar to their
analysis we assume flat priors inside the intervals $\om \in
[0.01;1.0]$ and $\sig \in [0.4;1.4]$, and zero prior otherwise.

\subsection{The covariance of the COSEBIs}
In Fig.\ts\ref{fig:correlation} we have plotted the correlation matrix
of the COSEBIs, defined as
\be
r_{mn}={C^{\rm E}_{mn}\over \sqrt{C^{\rm E}_{mm}\,C^{\rm E}_{nn}}}\;,
\elabel{correlrmn}
\ee
for several values of $m$, using both linear (upper panel) and
logarithmic (lower panel) weight functions. The value of $m$ can be
identified as the point where $r_{mn} = 1$. For the
linear weight functions, we see that the correlation matrix declines
quickly for $n\ne m$, reaches a (negative) minimum at $n=m\pm 2$, and
essentially is zero for $|m-n|\ge 4$. Thus, the covariance matrix is
in essence a band matrix. For the logarithmic COSEBIs, the non-zero
correlations between the $E_n$ span a larger range in $|m-n|$. One
therefore expects that the inversion of the covariance matrix for a
given $n_{\rm max}$ is more difficult for the logarithmic COSEBIs than
for the linear ones. However, as we will show below, a smaller number
of logarithmic COSEBIs are needed to extract all the cosmological
information contained in the shear correlation functions, compared to
the linear COSEBIs.

\subsection{\llabel{FOMdisc}Figures of Merit: a short discussion}

In order to illustrate the information content one usually calculates
the so-called credible regions, inside of which the true set of
parameters is located with a probability of e.g. 68\%, 95\%, 99.9\%.
Instead of showing likelihood contours for all cases considered, we
use two different measures to quantify the size of these credible
regions, where each measure characterizes the information contents
through a single number.

The first measure, $q$, is calculated from the determinant of the
second-order moment of the posterior likelihood $p(\vpi|\vec d)$, 
\be
\mathcal Q_{ij} \equiv \int \tn d^2 \vpi \, 
p(\vpi|\vec d) \; (\pi_i-\pi_i^{\mr f})(\pi_j -\pi_j^{\mr f})\,,
\elabel{quadrups}
\ee
where $\pi_i$ are the parameters of the model, and $\pi_i^{\mr f}$ are
the parameters of the fiducial model (here, $i=1,2$, corresponding to
$\om$ and $\sig$). We quantify the size  of the credible region by the 
square root of the determinant of $\mathcal Q$,
\be
q= \sqrt{|\mathcal Q_{ij}|} = \sqrt{\mathcal Q_{11} \mathcal Q_{22} 
- \mathcal Q_{12}^2}.
\elabel{detquadrups}
\ee 
Smaller credible regions in parameter space correspond to smaller
values of $q$. In this paper, all $q$'s are given in units of $10^{-4}$.

\begin{figure}
\includegraphics[width=9cm]{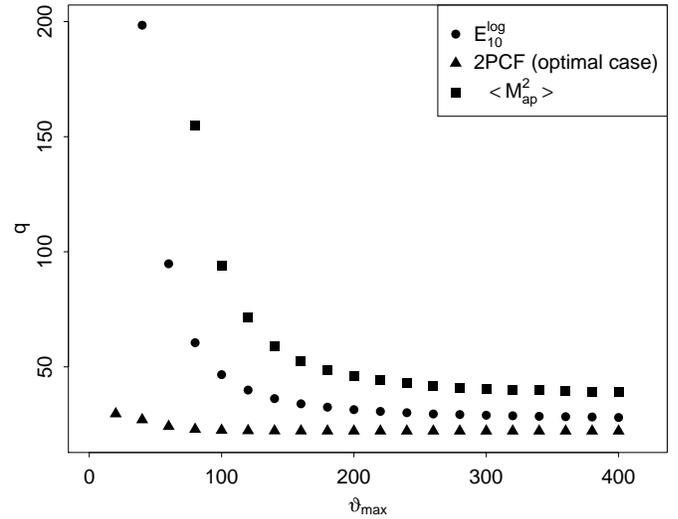}
\caption{The $q$ of the COSEBIs as a function
  $\vartheta_\mr{max}$, for $\tmin=1'$. The COSEBIs are calculated from
  $T^\mr{log}_n$, where $n$ ranges from $1 - 10$}
\label{fig:q_vartheta}
\end{figure}

Our second figure of merit is obtained from the Fisher information
matrix \citep{tth97}
\be
 F_{ij}=\frac{1}{2} \mr{tr} \left[ \tens{C}^{-1} \tens{C}_{,i} 
\tens{C}^{-1} \tens{C}_{, j}+ \tens{C}^{-1} \tens{M}_{ij} \right] \, ;
\elabel{fishergauss}
\ee
where subscripts separated by a comma denote partial derivatives with
respect to $\pi_i$, and $\tens{M}_{ij}
= \vec{\mr E}_{,i} \vec{\mr E^{\mr t}}_{,j} + \vec{ \mr E}_{,j}
\vec{\mr E^{\mr t}}_{,i}$, where $\vec{\mr E}$ is the $n_{\rm max}$-dimensional
vector of the first $n_{\rm max}$ $E_n$'s. The $n_{\rm max}\times
n_{\rm max}$-dimensional covariance matrix
$\tens{C}$ has the elements $C^{\rm E}_{mn}$, as given in
Eq.\ts(\ref{eq:CEmn}). We consider a constant covariance in
parameter space, so that the first term of Eq.\ts(\ref{eq:fishergauss})
vanishes. Since the Fisher matrix is the Hessian of the (negative of
the) log-likelihood function at its maximum, its elements describe the
size and shape of ellipses of constant likelihood near the maximum. If
the likelihood was strictly Gaussian, the Fisher matrix would
completely describe its functional form. We define our second figure
of merit $f$ as 
\be
f= {1\over \sqrt{\det(\tens{F})}} \,. 
\elabel{fdefinition}
\ee
For a better comparison with $q$ we chose to modify the more commonly
used figure of merit definition \citep[see e.g.,][]{abc06} -- we
consider the area of the error 
ellipse itself, not its inverse. Similar to $q$, $f$ is
given in units of $10^{-4}$.  With the definition
(\ref{eq:fdefinition}), $q$ and $f$ give the same result if (1) the
likelihood in the parameter space considered is Gaussian and (2) if
the likelihood outside the region where we set a flat prior is
negligible. We note that $f$ and $q$ can be significantly different if
these two assumptions are not satisfied. Then, the Gaussian defined by
the Fisher matrix is only a useful approximation close to the fiducial
model, and the resulting values of $f$ can be rather bad
approximations for $q$. In contrast,
$q$ is sensitive to parameter regions far from the fiducial model and
we therefore consider $q$ as the more useful measure for the
information contents. In order to give an impression of the meaning of
different $q$ and $f$ we show a sample of likelihood contours in
Fig.\ts\ref{fig:contour_EB} -- it is obvious that the likelihood
function in our case is far from Gaussian.

\begin{figure*}
\includegraphics[width=9cm]{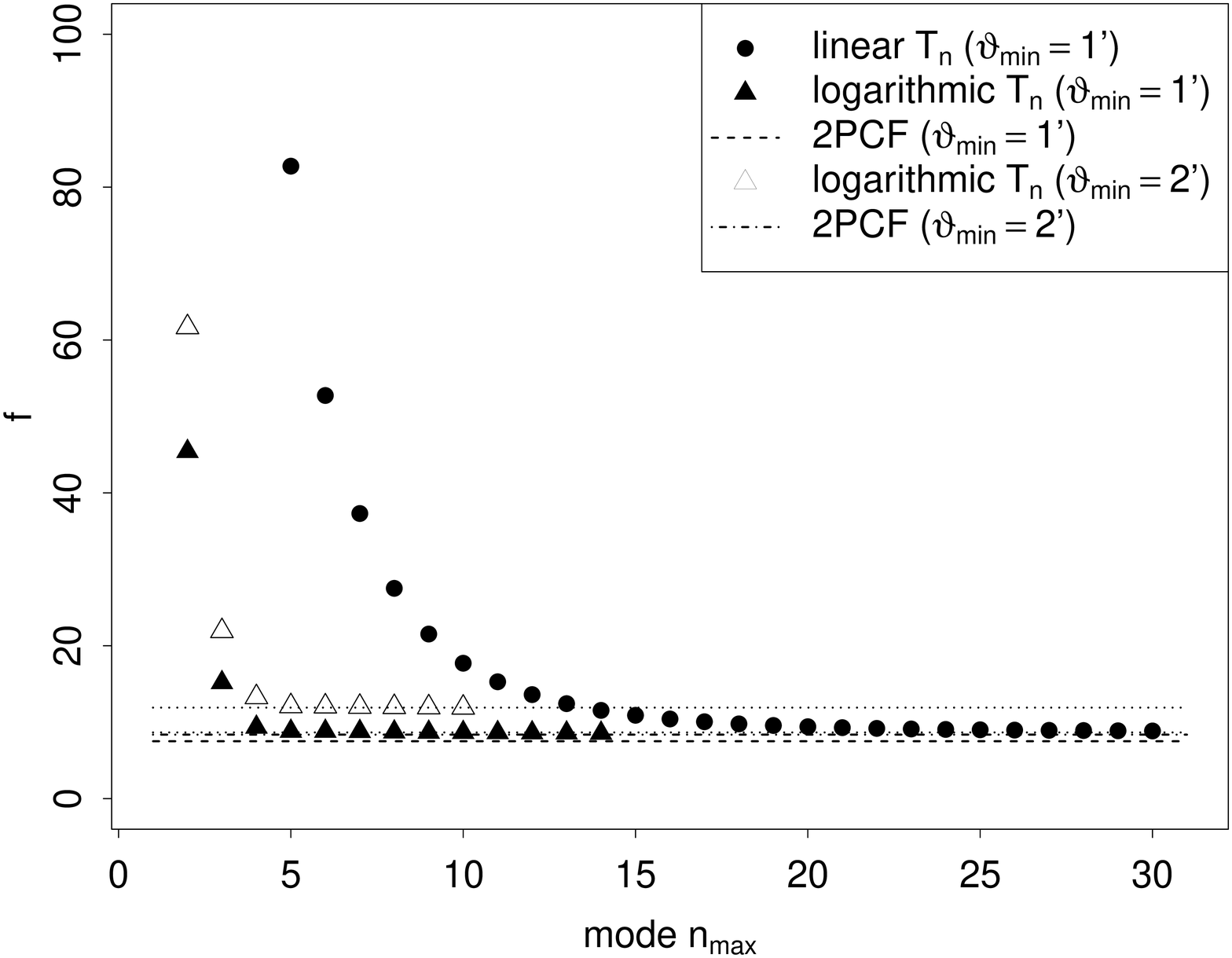}
\includegraphics[width=9cm]{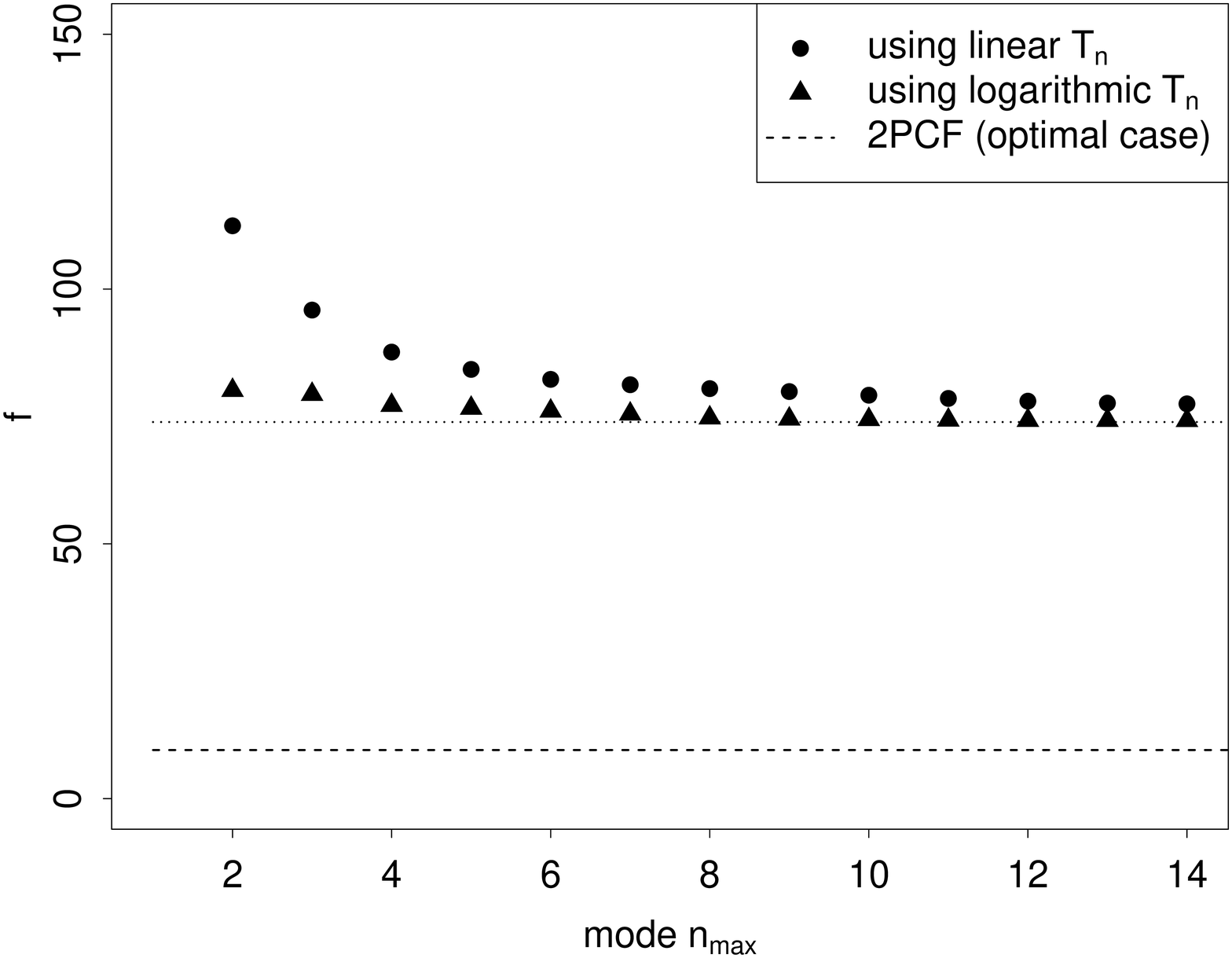}
\caption{The value of $f$ -- see Eq.\ts(\ref{eq:fdefinition}) -- for the
  case of linear (\textit{circles}) or logarithmic
  (\textit{triangles}) $T_n$-functions as a function of the maximum mode
  $n_{\rm max}$ which was included in the likelihood analysis. 
The results in the
  \textit{left} (\textit{right}) panel correspond to
  $\vartheta_\mr{max}=400'$ ($\vartheta_\mr{max}=20'$), and the filled
  symbols are calculated for $\tmin=1'$; in the left panel, we also
  plot corresponding results for $\tmin=2'$, indicated by the open
  triangles.  The dashed (dash-dotted) line represents the optimal $f$
  for $\tmin=1'$ ($\tmin=2'$), obtained when using the 2PCFs
  directly. The dotted lines shows the asymptotic value of $f$
  achieved for large $n_{\rm max}$ }
\label{fig:EBFOM}
\end{figure*}

\subsection{\llabel{sc:likeresults}Results of the likelihood analysis}
Figure \ref{fig:EBq} shows the values of $q$ for the case of
$\vartheta_\mr{max}=400'$ (left panel) and for
$\vartheta_\mr{max}=20'$ (right panel). The triangles correspond
to the COSEBIs from $T_{n}^\mr{log}$, whereas the circles correspond
to the COSEBIs calculated using the linear $T_{n}$. For comparison we show the
information content of the 2PCF (dashed line), which serves as an
upper limit on the information content of any second-order
E/B-decomposing measure -- since the 2PCFs contain all
information from second-order shear measurements and the COSEBIs are
derived from them \citep{eks08}.

For $\vartheta_\mr{max}=400'$ the minimum value of $q$ obtainable from
the COSEBIs -- and thus the available information on the two
cosmological parameters considered -- is extremely close to that
obtained from the 2PCFs. The logarithmic $E_n$ reach this threshold
already for $n_{\rm max}=5$, whereas the linear $E_n$ saturate around
$n_{\rm max}=25$, indicating that the logarithmic modes capture the
bulk of cosmological information in significantly fewer data points
compared to the linear case. This property can be particularly
important in higher-dimensional parameter spaces, where data
compression and computing time become important.

The COSEBIs for $\vartheta_\mr{max}=20'$ saturate much earlier; the
information content of $E_n$ is hardly increased when going beyond
$n=4$ ($n=3$ for the logarithmic weight functions). More important,
however, is the large difference between the saturation limit of the
COSEBIs and the corresponding information content of the 2PCFs (which
is also seen in the likelihood contours of
Fig.\ts\ref{fig:contour_EB}). Obviously, the choice of
$\vartheta_\mr{max}$ has a significant impact on the information
content, and on the relative information contained in the COSEBIs and
the 2PCFs. 

This latter difference is not due to a deficiency of the COSEBIs --
since they form a complete set of E-/B-mode measures, they contain all
the information that can uniquely be split into the two modes. If,
however, one \emph{assumes} that the shear field has no B-mode
contribution, and thus using of the full 2PCFs obviously yields
tighter parameter constraints. But, this assumption will hardly be
justifiable in any of the forthcoming surveys. The fact that the
measured B-modes are compatible with zero within the error bars in a
data set is \emph{not} a justification -- since any realistic survey
may contain B-modes which cannot be identified as such, for example a
uniform shear field which can either be E- or B-mode. Therefore, the
loss of information due to a clean mode separation is inevitable, but
a small price to pay relative to a potential bias of results due to
undetected B-modes. Fortunately, for surveys which allow shear
correlation measurements on large angular scales, this information
loss is seen to be almost negligible.

We analyse this more closely in Fig.\ts\ref{fig:q_vartheta}, where we
show $q$ as a function of $\vartheta_\mr{max}$; here we use
logarithmic weight functions with 10 modes, i.e., where the asymptotic
limit is well achieved.  The amount of information increases
significantly when going from $20'$ to $100'$ and becomes almost
constant when going to larger $\vartheta_\mr{max}$. This behavior, of
course, depends on the parameter space considered; for $\om$-$\sig$ it
can be understood from the functional behavior of the power
spectrum. For small $\ell$, it is almost fully degenerate in these two
parameters, hence going to larger angular scales does not yield
significantly more information -- this will be different for other
parameter combinations. One also sees that the difference in
information content between the COSEBIs and the 2PCFs decreases for
larger $\tmax$ -- the larger $\tmax$, the smaller is the contribution
of modes to the 2PCFs which can not be uniquely decomposed into
E/B-modes. Furthermore, we have plotted the corresponding values of
$q$ for the aperture dispersion $\ave{M_{\rm ap}^2(\theta)}$, where
$\theta=\tmax/2$ is the aperture radius which is calculated from the
shear 2PCFs for $\vt\le \tmax$. Values for $\ave{M_{\rm
    ap}^2(\theta)}$ are calculated and plotted only for $\theta\ge
40'$, to limit the bias caused by the lack of measured correlation
functions for $\vt < \tmin$ \citep[see][]{kse06} to $< 5\%$.  We see
that its information content is smaller than that of the COSEBIs, as
must be the case, owing to the completeness of the latter.

Figure \ref{fig:EBFOM} shows a similar analysis based on $f$. The
results confirm our foregoing findings. Similar to the case
of $q$, the Fisher matrix analysis shows that the logarithmic $E_n$
reach the saturation limit much earlier than the linear $E_n$ and
again, the saturation limit for $\vartheta_\mr{max}=400'$ is closer to
the optimal information content than for $\vartheta_\mr{max}=20'$. 

\begin{table}
\caption{Values of $q$ and $f$ as obtained by considering the full
  2PCFs, and by using the COSEBIs $E^\mr{lin}$, and $E^\mr{log}$}
\label{table1}
\centering
\renewcommand{\arraystretch}{1.2}
\label{tab:q-f-values}
\begin{tabular}{l l l l l}\hline \hline
Measure &$\vt_\mr{min}$ & $\vt_\mr{max}$& $q$ &$f$  \\ \hline
2PCF &1'& 400'& 22.10& 7.51 \\
$E^\mr{lin}_{30}$&1'& 400' & 28.68& 8.85 \\
$E^\mr{log}_{10}$&1'& 400' & 27.04& 8.66 \\
\\
2PCF&1'& 20' & 31.28& 9.54 \\
$E^\mr{lin}_ {15}$&1'& 20' &422.79 & 74.21 \\
$E^\mr{log}_{10}$&1'& 20'  &418.09& 77.35 \\
\\
2PCF&2'& 400' & 25.46& 8.37 \\
$E^\mr{log}_{10}$&2'& 400' &39.65 & 11.91\\
\hline \hline
\end{tabular}
\end{table}

Note that in Figs.\ts\ref{fig:EBq} and \ref{fig:EBFOM} we choose a
similar scale for the vertical axis in the right and the left panels
to enable for an easier comparison between the different cases of
$\vartheta_\mr{max}$. We point out the good agreement between the
saturation limits of $E_n$ and $E^{\rm log}_n$ in all cases, which
shows that our results are numerically robust. In
  Table\ts\ref{table1}, we have listed the values of $q$ and $f$ as
  shown in Figs.\ts\ref{fig:EBq} and \ref{fig:EBFOM} for the maximum
  number $n_{\rm max}$ of modes. The small difference between these
  values as obtained from the linear and logarithmic weight functions
  for the COSEBIs is due to the fact that for these values of $n_{\rm
    max}$, the linear ones have not yet reached their full asymptotic
  value.

The underlying reason why the formal loss of information of the
  COSEBIs, relative to the full 2PCFs, is larger for smaller $\tmax$
  is due to the filter functions that relates the 2PCFs to the power
  spectrum. This filter function is $J_0(x)$ for the case of
  $\xi_+(\vt)$, i.e. a function that tends towards $+1$ for small
  arguments. This implies that the correlation function $\xi_+(\vt)$
  is sensitive to long-range modes, i.e., modes of small $\ell$. In
  particular, this means that $\xi_+$ is also sensitive to the power
  spectrum for modes satisfying $\ell\le 2\pi/\tmax$, corresponding to
  scales which are in fact not probed by the 2PCFs directly -- and for
  which no E-/B-mode separation is possible from the data. The
  relative cosmological information content of the power spectrum in
  the ranges  $\ell\le 2\pi/\tmax$ and $2\pi/\tmax\lesssim
  \ell\lesssim 2\pi/\tmin$ decreases with increasing $\tmax$, which
  explains the difference in `relative information loss' in
  Figs.\ts\ref{fig:EBq} through \ref{fig:EBFOM}.

Up to now we have always chosen $\tmin=1'$. However, one may ask
  whether cosmic shear measurements down to this angular scale can be
  compared to sufficient accuracy with cosmological predictions, since
  at the corresponding length scales, baryonic physics can have a
  significant influence on the projected power spectrum. Of course,
  modeling the behavior of baryons in a cosmological simulation is
  much more difficult, and burdened with higher uncertainty, than dark
  matter-only simulations. \cite{Jing06} compared pure dark matter
  simulations with hydrodynamic simulations to conclude that for
  $\ell\sim 10^4$, corresponding to $\vt\sim 1'$, the predicted power
  spectra differ by about 10\% -- much more than the predicted
  statistical uncertainty of future cosmic shear surveys. 

  Fortunately, the largest effect of baryons on the total mass
  distribution seems to be a change of the halo concentration
  parameter as a function of halo mass \citep[][]{RuddZent2008}, in
  that baryons render halos more concentrated. If this is the case,
  then this effect can be calibrated from the weak lensing data
  themselves.  \cite{ZentnerRudd2008} studied such a self-calibration
  method for future surveys and concluded that the concentration--mass
  relation can be determined from the weak lensing data. In the
  framework of the halo model for the large-scale structure, the power
  spectrum can then be calculated using this modified
  concentration--mass relation, and fairly accurate model predictions
  can be made.

Dropping the small angular scales from future surveys implies
considerably weaker cosmological constraints. In the left panels of 
Figs.\ts\ref{fig:EBq} and \ref{fig:EBFOM}, we have plotted the values
of $q$ and $f$, respectively, for surveys with $\tmin=2'$. Independent
of whether the `optimal' constraints from the 2PCFs or the COSEBIs are
employed, the resulting constraints are weaker than for
$\tmin=1'$. Therefore, it is of considerable interest to improve the
accuracy of predictions for the matter power spectrum to small scales,
to make full use of the information contained in cosmic shear surveys
on small angular scales.

\section{Summary and discussion}
We have defined pure E- and B-mode cosmic shear measures from
correlation functions over a finite interval $\tmin\le \vt\le
\tmax$. These are complete orthonormal sets of such measures, implying
that they contain all cosmic shear information in the two-point
correlation functions which can be uniquely split into E- and
B-modes. For these COSEBIs, we have calculated their relation to the
underlying power spectrum and their covariance matrix.
Two different sets of COSEBIs have been explicitly constructed,
those with weight functions which are polynomials in the angular
scale, and those with polynomial weight functions in the logarithm of
the angular scale. For the former case, analytic expressions were
obtained for all orders, whereas in the logarithmic case, a linear
system of equations needs to be solved numerically.

\subsection{Advantages of the COSEBIs}
Comparing the COSEBIs with earlier cosmic shear measures, we point out
a number of advantages. First, using the correlation functions
themselves does not provide an E-/B-mode separation. The construction
of E-/B-mode correlation functions as described in \citep{cnp02}
requires knowledge of the correlation functions over an infinite
angular range, and is therefore not applicable in practice
(extrapolating to infinite separation using fiducial cosmological
models corresponds to `inventing data', and implicitly assumes that
there are no long-range B-modes). In fact, the generalization of pure
E-/B-mode correlation functions based on data over a finite angular
range has been derived here (see Sect.\ts\ref{sc:EB-decomp} and
Appendix\ts\ref{sc:EBmodeCFs}); however, we expect these to be of
limited use in practice.

Whereas the aperture mass dispersion \citep{swj98} provides a clean
separation into E- and B-modes \citep{cnp02,svm02}, it requires the
knowledge of the correlation function to arbitrarily small angular
separation. There are at least two aspects which render this
impractical: first, galaxy images need a minimum separation for their
shapes to be measurable. Second, on very small scales baryonic effects
will affect the power spectrum and render model predictions very
uncertain. The inevitable bias of the aperture mass dispersion
\citep{kse06} motivated the ring statistics \citep{sck07}. The latter removes
the bias, depends only on the correlation function over a finite
interval, and has potentially higher sensitivity to cosmological
parameters \citep[][FK10]{esk09}. However, the weight function of the
ring statistics is largely arbitrary.

The COSEBIs contain all available mode-separable information from the
correlation functions on a finite interval, and are therefore
guaranteed to provide highest sensitivity to cosmological
parameters. Furthermore, they form a discrete set of measures, whereas
the other cosmic shear statistics include a somewhat arbitrary grid of
variables, like the outer scale of the ring statistics: if the grid is
too coarse, information gets lost, whereas a finer grid renders the
measures largely redundant, implying large and significantly
non-diagonal covariances. In contrast, the discreteness of COSEBIs
leaves no freedom, and for the linear weight functions, the
covariances have a narrow band structure. The information clearly saturates
after a number of modes, and this number is surprisingly small for the
logarithmic weight function. Therefore, determining covariance
matrices from numerical simulations \citep[as was done for the COSMOS
analysis of][]{shj09} appears considerably simpler than for other
cosmic shear measurements, which is particularly true for an unbiased
estimate of their inverse \citep[see][for a discussion of this
point]{hss07}. Based on these properties of the COSEBIs, we would like
to advertise them as the method of choice for future cosmic shear
analyses.

\subsection{Generalizations}
In case photometric redshift information of the lensed galaxies is
available and several source populations can be defined based on their
redshift estimates, the COSEBIs can be generalized to a tomographic
version. Furthermore, under the same assumption, intrinsic alignment
effects between the tidal gravitational field and the intrinsic galaxy
orientation \citep[e.g.,][]{ckb01,cnp01,jin02,his04} can be filtered
out by properly choosing redshift-dependent weight functions, such as
to avoid physically close pairs of galaxies \citep{kis02,kis03,heh03}
or make use of the specific redshift dependence of the shear-intrinsic
alignments \citep{brk07,jos08,jos09}, possibly in combination with
other data \citep{job09}. We expect that these generalizations of the
COSEBIs provide no real difficulties.

It would be desirable to obtain a similar measure for third-order
cosmic shear statistics, i.e., one that provides clear E-/B-mode
separations from three-point correlation functions measured over a
finite interval. Up to now, the aperture statistics is the only known
such measure \citep{jbj04,skl05}; however, similar to the case of the
aperture dispersion, third-order aperture statistics requires the
correlation functions to be measured down to arbitrarily small
separations. A generalization of the COSEBIs to third order seems
challening -- not only because of the higher number of independent
variables (the three-point correlation functions depend on three
variables) and the larger number of modes (one pure E-mode, one mixed
E/B-mode, and two further modes which are not invariant under parity
transformation), but also because of the more complicated relation
between correlation functions and the bispectra \citep{skl05}. Thus,
even the analogue of the starting point of the current paper --
Eqs.\ts(\ref{eq:EBmodes}) and (\ref{eq:Tplusminus}) -- is not yet
known for the third-order case.

\begin{acknowledgements}
  We thank Liping Fu and Martin Kilbinger for interesting discussions
  on E-/B-mode separations which triggered this study, and an
  anonymous referee for constructive suggestions. We thank Marika Asgari
  for checking some of the numerical results presented here. This work was
  supported by the Deutsche Forschungsgemeinschaft within the
  Transregional Research Center TR33 `The Dark Universe' and the
  Priority Programme 1177 `Galaxy Evolution' under the project SCHN
  342/9.

\end{acknowledgements}

\begin{appendix}
\section{Calculation of the COSEBIs: Numerical problems 
and solutions}  
\llabel{appendix1}
Several numerical issues arose during the implementation of the
calculations of the COSEBIs, especially in the context of their
covariance. As these issues are crucial for obtaining the correct
values of $q$ and $f$, we outline them in greater detail.  We employ
the QAG adaptive integration routine from the GNU Scientific
Library\footnote{http://www.gnu.org/software/gsl/} and obtain the
$E_n$ using two different methods. First, we calculate them from the
set of 2PCFs according to Eq.\ts(\ref{eq:EBmodes}) and second, we
check for consistency by calculating $E_n$ directly from the $P_\mr E$
according to Eq.\ts(\ref{eq:EBfromP}). The first method cleanly
separates E- and B-modes, giving a B-mode residual due to numerical
uncertainties which is 8 to 5 orders of magnitudes lower than the
E-mode, depending on the scales considered and whether one uses
$T_{n}$ or $T_{n}^\mr{log}$. Both methods yield results in perfect
agreement, hence we are confident that there are no numerical problems
in either of them.

When using a binned version of the 2PCF instead, we find a
non-negligible deviation when using too few angular bins. The
number of bins, above which the E/B-decomposition becomes stable,
depends on the mode $E_n$, the maximum scale $\tmax$ of the 2PCF, and
whether one uses $T_{n}$ or $T_{n}^\mr{log}$. This 
should be checked carefully before applying the method to an actual
data set. As an example, we found that for linear $T_{n}$ and
$\vt_\mr{max}=400'$, one needs $\sim 10^5$ bins to calculate $E_{30}$
properly and to have an accurate mode separation.

The calculation of the covariance $C^{\rm E}_{mn}$ is numerically more
challenging than that of the data vectors. Again, we use two
approaches and calculate $\tens{C}^{\rm E}$ from $P_\mr E$ using
Eq.\ts(\ref{eq:CEmn}), and from the 2PCF covariance using
Eq.\ts(\ref{eq:CE2PCF}). Both methods have their difficulties and
need to be checked carefully for consistency before using the
covariance in the likelihood analysis.

The power spectrum approach using Eq.\ts(\ref{eq:EBfromP}) involves
the calculation of a one-dimensional integral over $P_{\rm E}$
multiplied by two filter functions $W_{\pm}$. As can be seen from
Figs.\ts\ref{fig:W_lin} and \ref{fig:Wlog}, these filter functions are
strongly oscillating, which becomes worse for large $\ell$ and higher
modes $n$. We use a stepwise integration to calculate the integral and
truncate the integral once the ratio of ``new contribution in step $i$
/ integral calculated until $(i-1)$'' drops below a certain
threshold. We vary the width of the steps as well as the truncation
threshold; however, we find that the integration becomes
inaccurate when going to higher modes $n$.

For calculation of $C^{\rm E}_{mn}$ from the covariance of the 2PCFs we
find that it is too time-consuming to calculate the 2PCF covariance
for every sampling point of the integration routine
separately. Instead, we calculate the 2PCF covariance for a specific
binning and interpolate the values during the integration. We use a
linear binning in the 2PCF covariance for the linear weight function
and a logarithmic binning for the case of $T_n^\mr{log}$. In addition,
we check how strongly the number of bins influences the accuracy of the
integral, finding that we can calculate $\tens{C}^{\rm E}$ properly if we
choose at least $1000 \times 1000$ bins in the 2PCF covariance. The
final $\tens{C}^{\rm E}$ must be symmetric, positive definite, and not
ill-conditioned, as already small deviations from these requirements
can bias the information content measures $q$ and $f$.

\section{\llabel{SN-maxi}S/N maximization}
From a complete set of functions $T_{+n}$ obeying the constraints
(\ref{eq:conditions}) for given $\tmin$ and $\tmax$, we can find a
weight function $T_+(\vt)$ which maximizes the signal-to-noise of the
E-mode. This problem was also considered by FK10. In this case, we can
write 
\be
T_+(\vt)=\sum_{n=1}^N a_n\,T_{+n}(\vt)\;,
\elabel{PS11}
\ee
which satisfies the integral constraints (\ref{eq:conditions}) for any
choice of the $a_n$. Then the E-mode signal is, in the absence of
B-modes, 
\be
E=\int_{\tmin}^{\tmax}\d\vt\;\vt\,T_+(\vt)\,\xi_+(\vt)
=\sum_{n=1}^N a_n\,E_n\;.
\elabel{PS12}
\ee
The noise ${\rm N}$ of $E$ is obtained through the covariance of the
$E_n$, 
\be
{\rm N}^2=\ave{E^2}-\ave{E}^2
=\sum_{m,n=1}^N a_m a_n C^{\rm E}_{mn}\;,
\elabel{PS14}
\ee
yielding as signal-to-noise ratio
\be
{\rm S\over N}={\sum_n a_n E_n \over \sqrt{\sum_{m,n} a_m a_n C^{\rm E}_{mn}}} \;.
\elabel{PS16}
\ee
To obtain a maximum of $\rm S/N$ with respect to the coefficients
$a_n$, we differentiate the foregoing expression with respect to a
coefficient $a_k$,
\bea 
{\partial \over \partial a_k}{\rm S\over N}&=& {E_k\over
  \sqrt{\sum_{m,n} a_m a_n C^{\rm E}_{mn}}} \nonumber \\
&& -{\sum_n a_n E_n
\over 2 {\rm N}^{3}}
\sum_{m,n}\rund{\delta_{mk}a_n C^{\rm E}_{mn}+\delta_{nk}a_m C^{\rm E}_{mn} } 
\nonumber \\
&=& {\rm
  N}^{-3}\eck{{\rm N}^2 E_k-\rund{\sum_n a_n E_n}\rund{\sum_n
    C^{\rm E}_{kn}a_n}} \;.
\elabel{PS17}
\eea
Setting this derivative to zero results in
\be
E_k = {\sum_n a_n E_n \over \sum_{m,n} a_m a_n C^{\rm E}_{mn}}
\sum_n C^{\rm E}_{kn} a_n\;.
\elabel{PS18}
\ee
From this equation we see that the overall amplitude of the $a_n$
cannot be determined, i.e., if the $a_n$ are a solution, then $\lambda
a_n$ solve the equation as well. Noting that the first term on the
r.h.s. of Eq. (\ref{eq:PS18}) does not depend on $k$, a solution is
obtained as 
\be
a_k=\sum_n \rund{C^{\rm E}}^{-1}_{kn} E_n\;,
\ee
as can be also verified by inserting this into Eq. (\ref{eq:PS18}).  Thus,
if the function $T_+$ is expanded into a set of functions which all
satisfy the constraints (\ref{eq:conditions}), the signal-to-noise
maximization can be done analytically. If different sets of functions
are used for constructing the $T_+$ maximizing the S/N, the resulting
function should be the same in the limit $N\to \infty$; however,
different sets of functions may require different $N$ before the
asymptotic limit is reached.

\section{\llabel{EBmodeCFs}Pure E-/B-mode correlation functions}
We will now explore how the pure-mode correlation functions 
introduced in Eq.\ts(\ref{eq:EBpureFCs})
are
related to the original $\xi_\pm$. For this, we use Eq.\ts(\ref{eq:EBmodes})
in the definition (\ref{eq:EBpureFCs}) to obtain
\bea
\xi_\pm^{\rm E,B}(\vt)&=&
\int_\tmin^\tmax {\d \vp\;\vp\over \vt\,\Delta\vt}
\Bigl[ \xi_+(\vp)\sum_{n=1}^\infty T_{\pm n}(\vt) T_{+n}(\vp) \nonumber\\
&& +\mu\,\xi_-(\vp)\sum_{n=1}^\infty T_{\pm n}(\vt) T_{-n}(\vp) \Bigr]
\elabel{xiEB}
 \\
&=&
\int_\tmin^\tmax {\d \vp\;\vp\over \vt\,\Delta\vt}
\eck{\xi_+(\vp) S_{\pm +}(\vt,\vp)+\mu\,\xi_-(\vp) S_{\pm -}(\vt,\vp)}
\;, \nonumber
\eea
where $\mu=+1$ for E-modes, $\mu=-1$ for the B-modes, and where we
defined the functions $S_{\pm \pm}(\vt,\vp)$ in the last step. These
functions are calculated next, by noting
that the normalized Legendre polynomials $p_n(x)$ as defined in
Eq.\ts(\ref{eq:tplusn}) are orthonormal,
\[
\int_{-1}^1\d x\;p_n(x)\,p_m(x)=\delta_{mn}\;,
\]
form a complete set of functions on the interval $[-1,1]$, and
therefore obey
\be
\sum_{n=0}^\infty p_n(x)\,p_n(y)=\delta_{\rm D}(x-y)\;.
\ee
Noting that we have chosen in Sect. \ref{sc:polynomials} $t_n(x)=p_{n+1}(x)$
for $n\ge 3$, we find that
\bea
s_{++}(x,y) 
&=&\sum_{n=1}^\infty t_{+n}(x)\,t_{+n}(y)\nonumber \\
&=& \sum_{n=4}^\infty p_n(x)\,p_n(y)+\sum_{n=1}^2 t_{+n}(x)\,t_{+n}(y)
 \nonumber\\
&=& \delta_{\rm D}(x-y)-\! \sum_{n=0}^3 p_n(x)\,p_n(y)
       +\! \sum_{n=1}^2 t_{+n}(x)\,t_{+n}(y) \nonumber\\
&=:&\!\!  \delta_{\rm D}(x-y) -F_{++}(x,y)\;, 
\eea
where in the final step we have defined the function $F_{++}(x,y)$, which
is obviously symmetric in its arguments. 
The explicit expression for
it reads
\bea
F_{++}\!\!\!\!\!&&\!\!\!\!\!(x,y)={5 (1+B x) (1+B y)\over 8(175+35 B^2 + 45
  B^4 +B^6)}\nonumber \\ 
&\times&
\Big(
140(1+3 x y) 
+70 B (3 x y -5)(x+y)\nonumber \\
&+&7B^2[39+20 x y -25(x^2+y^2)+15 x^2
y^2]
\\
&+& 14 B^3 (5 x y-3)(x+y)
+ B^4[15-21(x^2 + y^2)+35 x^2 y^2] \Big) \;.\nonumber
\eea
We can now calculate the other sums in Eqs. (\ref{eq:xiEB}), making use of
Eq. (\ref{eq:tminusx}) written in the form
\be
t_{-n}(x)=t_{+n}(x)+\int_{-1}^x\d z\;t_{+n}(z)\,{\cal G}(z,x)\;,
\ee
with
\be
{\cal G}(z,x)={4 B\over (1+B x)^2}\eck{1+B z-{3(1+B z)^3\over (1+
    Bx)^2}}\;.
\ee
This then yields
\bea
s_{+-}(x,y)&=&\sum_{n=1}^\infty t_{+n}(x)\,t_{-n}(y) \nonumber \\
&=& s_{++}(x,y)+\int_{-1}^y\d z\; s_{++}(x,z)\,{\cal G}(z,y)\\
&=&
\delta_{\rm D}(x-y) -  F_{++}(x,y) 
 + {\rm H}(y-x)\,{\cal G}(x,y)
-V(x,y)\;,\nonumber
\eea
where
\be
V(x,y)=\int_{-1}^y\d z\;F_{++}(x,z)\,{\cal G}(z,y)\;.
\ee
Owing to symmetry,
\be 
s_{-+}(x,y)=\sum_{n=1}^\infty t_{-n}(x)\,t_{+n}(y)
=s_{+-}(y,x)\;,
\ee
and
\bea
s_{--}(x,y)&=&\sum_{n=1}^\infty t_{-n}(x)\,t_{-n}(y)\nonumber\\
&=& s_{++}(x,y)+{\rm H}(y-x){\cal G}(x,y) \\
&+& {\rm H}(x-y){\cal G}(y,x)
-V(x,y)-V(y,x)+W(x,y)\;,\nonumber
\eea
where the symmetric function $W$ is defined as
\bea
W(x,y)&=&\int_{-1}^{{\rm min}(x,y)}\d z\;{\cal G}(z,x)\,{\cal
  G}(z,y)
\nonumber \\
&-&\int_{-1}^x\d z\int_{-1}^y \d z'\;F_{++}(z,z')\,
{\cal G}(z,x)\,{\cal G}(z',y)\;.
\eea
Thus, we find for the $S_{\pm\pm}(\vt,\vp)$ in turn, using
$x=2(\vt-\bar\vt)/\Delta\vt$ and $y=2(\vp-\bar\vt)/\Delta\vt$:
\bea
S_{++}(\vt,\vp)&=&{\Delta\vt\over 2}\,\delta_{\rm
  D}(\vt-\vp)-F_{++}(x,y)\;, \nonumber \\
S_{+-}(\vt,\vp)&=&{\Delta\vt\over 2}\,\delta_{\rm
  D}(\vt-\vp)-F_{+-}(x,y)
+ {\rm H}(\vp-\vt)\,{\cal G}(x,y)\;,
\nonumber \\
S_{-+}(\vt,\vp)&=&S_{+-}(\vp,\vt)\;,\\
S_{--}(\vt,\vp)&=&{\Delta\vt\over 2}\,\delta_{\rm
  D}(\vt-\vp)+{\rm H}(\vp-\vt)\,{\cal G}(x,y)\nonumber \\
&+&{\rm H}(\vt-\vp)\,{\cal G}(y,x)
-F_{--}(x,y)\;, \nonumber 
\eea
with $F_{+-}(x,y)=F_{++}(x,y)+V(x,y)$,
$F_{--}(x,y)=F_{++}(x,y)+V(x,y)+V(y,x)-W(x,y)$. 
We finally obtain for the pure mode correlation functions
\bea
\xi_+^{\rm E,B}(\vt)&=&{\xi_+(\vt)+\mu\xi_-(\vt)\over 2} \nonumber \\
&-& \int_\tmin^\tmax {\d \vp\;\vp\over \vt\,\Delta\vt}
\Bigl[ \xi_+(\vp)\,F_{++}\rund{ {2(\vt-\bar\vt)\over
      \Delta\vt},{2(\vp-\bar\vt)\over \Delta\vt}}\nonumber \\
&+&\mu\,\xi_-(\vp) \,F_{+-}\rund{ {2(\vt-\bar\vt)\over
      \Delta\vt},{2(\vp-\bar\vt)\over \Delta\vt}} \Bigr] \nonumber \\
&+&\mu\int_\vt^\tmax {\d \vp\;\vp\over \vt\,\Delta\vt}\; 
\xi_-(\vp)\,{\cal G}\rund{ {2(\vt-\bar\vt)\over
      \Delta\vt},{2(\vp-\bar\vt)\over \Delta\vt}}\nonumber \\
\xi_-^{\rm E,B}(\vt)&=&{\xi_+(\vt)+\mu\xi_-(\vt)\over 2} \nonumber \\
&+& \!\!\int_\tmin^\vt \! {\d \vp\;\vp\over \vt\,\Delta\vt} \;
{\cal G}\rund{{2(\vp-\bar\vt)\over \Delta\vt},{2(\vt-\bar\vt)\over
      \Delta\vt}} 
\eck{\xi_+(\vp)+\mu\,\xi_-(\vp)} \nonumber \\
&+&\mu\int_\vt^\tmax {\d \vp\;\vp\over \vt\,\Delta\vt}\;
\xi_-(\vp)\,{\cal G}\rund{ {2(\vt-\bar\vt)\over
      \Delta\vt},{2(\vp-\bar\vt)\over \Delta\vt}}\nonumber \\
&-& \int_\tmin^\tmax {\d \vp\;\vp\over \vt\,\Delta\vt}\,
\Bigl[\xi_+(\vp)\,F_{+-}\rund{{2(\vp-\bar\vt)\over
      \Delta\vt},{2(\vt-\bar\vt)\over \Delta\vt}}\nonumber \\
&+& \mu\,\xi_-(\vp)\,F_{--}\rund{ {2(\vt-\bar\vt)\over
      \Delta\vt},{2(\vp-\bar\vt)\over \Delta\vt}}\Bigr] \;.
\eea
Hence, pure mode correlation functions can be obtained from the
observed correlation functions over a finite interval. However, we
believe that these pure mode correlation functions are of little
practical use, since for a quantitative analysis of cosmic shear
surveys the COSEBIs contain all relevant information.
\end{appendix}


\end{document}